\documentclass[12pt,a4wide]{article}
\usepackage{afterpage}
\usepackage{epsfig}
\usepackage[a4paper]{geometry}
\usepackage{graphicx}	
\newcommand\ignore[1]{}			
\usepackage{amssymb}
\usepackage{amsmath}
\usepackage{amssymb}
\usepackage{amsthm}
\usepackage{graphicx}
\usepackage[urlcolor=blue,colorlinks=true,linkcolor  = black]{hyperref}
\usepackage{placeins}
\usepackage{tocloft}
\usepackage{slashed}
\usepackage{float}
\usepackage{fullpage}
\usepackage{braket}
\usepackage{url}

\usepackage{tabularx}

\usepackage{cite}
\usepackage{subcaption}
\usepackage{upgreek}
\usepackage{outlines}
\usepackage{braket}
\usepackage{appendix}
\def\0{{(0)}}
\def\1{{(1)}}
\usepackage{xspace}
\usepackage{ulem}

\usepackage{mathrsfs} 
\usepackage{comment}
\usepackage{hyperref}
\usepackage{physics}

\usepackage{bm}

\usepackage{caption}
\usepackage{subcaption}

\def\ccccend{\end{array}\right)}

\usepackage{tikz}
\usepackage{pgfplots}
\usetikzlibrary{matrix,arrows,decorations.pathmorphing}
\usetikzlibrary{patterns}
\usetikzlibrary{shapes.misc}
\usetikzlibrary{trees}
\usetikzlibrary{decorations.pathmorphing}
\usetikzlibrary{shapes.geometric}
\usetikzlibrary{positioning}
\usetikzlibrary{decorations.markings}
\usetikzlibrary{positioning,arrows}

\usetikzlibrary{decorations.pathreplacing}
\usetikzlibrary{decorations.pathmorphing}
\usetikzlibrary{shapes}
\usetikzlibrary{arrows.meta}
\usetikzlibrary{plotmarks}
\usetikzlibrary{decorations.markings}
\tikzset{
particle/.style={thin,draw=black, postaction={decorate},
decoration={markings,mark=at position .5 with {\arrow[black, line width=0.5mm]{stealth}}}},
gluon/.style={decorate, draw=black, decoration={coil,amplitude=4pt, segment length=5pt}},
photon/.style={decorate, decoration={snake}},
singularity/.style={decorate, draw=black, decoration=zigzag}
}

\usepackage[bottom]{footmisc}

\theoremstyle{definition}

\usepackage{amsthm}
\theoremstyle{theorem}

\usepackage{tcolorbox}

\begin{document}

\pagestyle{plain}

\makeatletter
\@addtoreset{equation}{section}
\makeatother
\renewcommand{\theequation}{\thesection.\arabic{equation}}
\pagestyle{empty}

\begin{center}
\phantom{a}\\
\vspace{0.8cm}
\scalebox{0.90}[0.90]{{\fontsize{24}{30} \bf{Universality in the Axiverse}} }
\end{center}

\vspace{0.4cm}
\begin{center}
\scalebox{0.95}[0.95]{{\fontsize{15}{30}\selectfont Junyi Cheng and Naomi Gendler}}
\end{center}

\begin{center}
\vspace{0.25 cm}
\textsl{Jefferson Physical Laboratory, Harvard University, Cambridge, MA 02138 USA}\\

	 \vspace{1.1cm}
	\normalsize{\bf Abstract} \\[8mm]
\end{center}
\begin{center}
	\begin{minipage}[h]{15.0cm}

Studies of axion effective theories in the type IIB Calabi-Yau landscape have revealed that hierarchies in cycle volumes drive correlations between axion physics and the number of axions in a given model. We analyze distributions of divisor volumes in toric hypersurface Calabi-Yau threefolds, and provide evidence that aspects of these distributions are universal across this landscape. Furthermore, we show that axion observables in this landscape can be approximated efficiently in terms of only divisor volumes. Finally, we propose a simple model for the spectrum of divisor volumes and use this model to reproduce results on axion masses and decay constants across the Kreuzer-Skarke axiverse.

	\end{minipage}
\end{center}
\vfill
\today
\newpage

\setcounter{page}{1}
\pagestyle{plain}
\renewcommand{\thefootnote}{\arabic{footnote}}
\setcounter{footnote}{0}
%
%
\setcounter{tocdepth}{3}

\tableofcontents

\newpage

\section{Introduction}

Enumerative analysis of the physics of weakly-coupled string compactifications has revealed a rich landscape of effective four-dimensional physics. The particle and force content of a given effective theory is dictated in a large part by the geometry of the internal compactification manifold. The past decade has seen incredible progress in computational methods for analyzing geometric and topological properties of Calabi-Yau threefolds \cite{Demirtas:2018akl,Demirtas:2020dbm, Demirtas:2023als, Gendler:2023ujl, Chandra:2023afu, Gendler:2022ztv, MacFadden:2024him, Dubey:2023dvu, Moritz:2023jdb}, rendering systematic studies of string compactifications on such manifolds tractable.

A large focus of such systematic studies has been on the physics of axion-like particles in the resulting effective theories. The presence of light axions is ubiquitious in this context \cite{Svrcek:2006yi, Conlon:2006tq, Arvanitaki:2009fg}, with masses and decay constants determined by the internal geometry of the compactification manifold. In particular, a lot of focus has been on studying compactifications of type IIB string theory on Calabi-Yau threefolds constructed as hypersurfaces in toric varieties defined by the comprehensive list of 4d reflexive polytopes in the Kreuzer-Skarke (KS) database \cite{Kreuzer:2000xy}.  In this context, strong correlations have been found between the number $N$ of axions in a given effective theory, and physical properties of the axions. Previous studies \cite{Mehta:2021pwf, Gendler:2023kjt, Gendler:2024adn, Demirtas:2021gsq, Demirtas:2018akl, Benabou:2025kgx, Halverson:2019cmy, Halverson:2019kna} have found  that as $N$ increases, axion-photon couplings increase, overall dark matter abundance decreases, and Peccei-Quinn quality increases. These phenomena can be traced to the following observation: as $N$ increases, hierarchies of four-cycle volumes in Calabi-Yau threefolds increase. The purpose of this work is to quantify this statement.

More specifically, the goal of this work is threefold: firstly, to show that axion effective theories arising from compactifications of type IIB string theory on Calabi-Yau threefolds can be efficiently approximated using only the data of the manifold's divisor volumes. Secondly, to provide a working model for the spectrum of divisor volumes in Calabi-Yau manifolds constructed as hypersurfaces in toric varieties. And finally, to use this model to reproduce broad-stroke axion statistics across the landscape of type IIB compactifications on such geometries.

One of the main results of this work is the observation that certain aspects of the divisor volume spectra across different toric hypersurface Calabi-Yau threefolds are universal and independent of the specific geometry. In particular, we analyze divisor volumes based on which kind of points they correspond to in the polytope defining the toric variety. A useful classification is by the smallest-dimension face in the polytope containing the point in question. Accordingly, we refer to these as \textit{vertex}, \textit{edge}, and \textit{face} divisors. We use statistical tests (the $1$-Wasserstein distance, also known as the Earth Mover's distance, and adjusted $r^2$ test) to show that distributions of appropriately normalized vertex, edge, and face divisors remain approximately constant as the geometry is varied. 

Finally, we propose a simple model that approximates the overall divisor volume distribution. The model consists of two components: the first, an analytic function that approximates the distribution of suitably normalized divisor volumes. The second is a regression on the mean divisor volume as a function of the Hodge number $h^{1,1}$ (or equivalently, $N$). Using this model, we are able to reproduce previous results on axion masses and decay constants, as well as on the quality of the QCD axion in this landscape. The utility of this model is that it can replace computationally expensive scans of effective axion theories with a nearly instantaneous approximation. Additionally, having a model for the spectra of divisor volumes of Calabi-Yau threefolds is a crucial step in understanding the underlying geometric structures that drive correlations in axion physics in the string theory landscape.

The paper is organized as follows: in \S\ref{sec:review}, we review the fundamentals of constructing toric hypersurface Calabi-Yau threefolds and the resulting geometries, as well as the axion effective theories resulting from compactifying type IIB string theory on these manifolds. In \S\ref{sec:axiondivs}, we explain our method of constructing an ensemble of Calabi-Yau threefolds from the Kreuzer-Skarke database, review the statistical tests that we utilize in this work, and use these tests to argue that divisor volumes constitute sufficient data for reconstructing axion effective theories. In \S\ref{sec:universal}, we provide evidence that normalized vertex, edge, and face divisor volumes come from separate underlying distributions that remain roughly constant across all geometries in our ensemble. In \S\ref{sec:model}, we present a simplified model for the overall divisor volume spectrum. We use this model to reproduce previous studies in the literature on axion observables. Finally, we conclude in \S\ref{sec:conclusions}.

\section{Review: Calabi-Yau threefolds and axion effective theories} \label{sec:review}

The context for our study is type IIB string theory compactified on Calabi-Yau threefold (orientifolds), constructed as hypersurfaces in toric varieties. Specifically, we are interested in the axion effective theories that result from such a compactification. In this section, we will review the fundamentals of the relevant geometry and subsequent axion physics. We begin in \S\ref{subsec:batyrev} with a summary of the Batyrev construction of Calabi-Yau threefolds as toric hypersurfaces. Then in \S\ref{subsec:cygeometry} we explain the relevant aspects of the geometry of Calabi-Yau threefolds and their moduli spaces. Finally, in \S\ref{subsec:axEFTs}, we explain how the physical observables of axions resulting from these compactifications are derived in terms of geometric quantities that we can compute.

\subsection{Batyrev construction of Calabi-Yau threefolds} \label{subsec:batyrev}

In this section, we will give a brief review of the Batyrev construction \cite{batyrev1993dualpolyhedramirrorsymmetry} of toric hypersurface Calabi-Yau threefolds in terms of four dimensional reflexive polytopes, before discussing a useful classification of divisors in this context. In particular, we will explain the geometric interpretation of the prime toric divisors at the level of the polytope, and discuss the qualitative features of their volumes.

The key concept in the construction \cite{batyrev1993dualpolyhedramirrorsymmetry} is the fact that there is an identification between fans constructed as unions of strongly convex rational cones and toric varieties. Moreover, the toric variety is compact if the toric fan is \textit{complete,} i.e. the cones span the entire space. The dimension of the toric variety is given by the dimension of the fan, and a hypersurface in that toric variety is complex codimension one in that space. If the toric variety is K\"ahler and its singularities are sufficiently mild, then the generic anti-canonical hypersurface in this four-dimensional toric variety defines a smooth, compact Calabi-Yau threefold.  In \cite{batyrev1993dualpolyhedramirrorsymmetry}, it was shown that the  toric fan for such a variety can be obtained by constructing an appropriate triangulation of a four-dimensional reflexive polytope, described below.

Let $\Delta^\circ$ be a four-dimensional lattice polytope, consisting of lattice points $\vec{p}_i$ (where we omit points interior to facets of $\Delta^\circ$). We take $\Delta^\circ$ to be \textit{reflexive}, i.e. its dual, $\Delta$, is also a lattice polytope. Let $\mathcal{T}$ be a triangulation of $\Delta^\circ$. $\mathcal{T}$ is a collection of simplices constructed from the points $\vec{p}_i$. We refer to the sets of various dimensional faces as $\mathcal{F}^0$ (vertices), $\mathcal{F}^1$ (edges), $\mathcal{F}^2$ (two-faces), and $\mathcal{F}^3$ (three-faces).

We furthermore restrict ourselves to the study of \textit{favorable} polytopes: a polytope is called favorable if every two-face containing points on its strict interior is dual to a one-face with no points on its strict interior in the dual lattice polytope. Favorable polytopes have the property that the toric varieties that they define have exactly $h^{1,1}+4$ primitive divisors, which descend to calibrated divisors in the Calabi-Yau threefold. We refer to such divisors as \textit{prime toric divisors}.

In order for the generic anti-canonical hypersurface in the toric variety defined by $\mathcal{T}$ to be a smooth, compact toric variety, $\mathcal{T}$ needs to be a \textit{fine, regular} and \textit{star} triangulation (FRST). The condition that $\mathcal{T}$ is fine means that every point $\vec{p}_i$ is a vertex of a simplex in $\mathcal{T}$, which ensures that the generic anticanonical hypersurface in the toric variety is smooth. The regularity condition means that $\mathcal{T}$ is a projection of the lower faces of a convex hull in a dimension one more than that of $\mathcal{T}$, and ensures that the toric variety is K\"ahler. The condition that $\mathcal{T}$ is star means that all simplices in $\mathcal{T}$ contain the origin as one of their vertices. This ensures that the toric fan defined by $\mathcal{T}$ is complete, and thus that the toric variety is compact. Finally, a Calabi-Yau threefold $X$ is constructed as a generic anticanonical hypersurface in the toric variety defined by $\mathcal{T}$.

The points $\vec{p}_i$ correspond to divisor classes in $X$. An important classification of these divisors is by their location in $\Delta^\circ$: we will refer to divisors corresponding to vertices as \textit{vertex divisors} ($\Sigma_i^{v}$), those corresponding to points strictly interior to edges as \textit{edge divisors} ($\Sigma_i^{e}$), and those corresponding to points strictly interior to two-faces as \textit{face divisors} ($\Sigma_i^{f}$). Points interior to three-faces correspond to divisors in the ambient toric variety that do not intersect $X$. 

Vertex, edge, and face divisors have qualitatively different properties. A divisor $D$ is called \textit{rigid} if it satisfies
\begin{align}
    \text{dim} \, H^0(D,\mathcal{O}_D) = 1, \ \ \text{dim} \,  H^i(D,\mathcal{O}_D) = 0 \  \forall i>0.
\end{align}
A divisor that is not rigid in the Calabi-Yau has a moduli space across which it sweeps out a dense subset in $X$ itself: therefore, shrinking this divisor to zero volume necessarily shrinks the entire Calabi-Yau. In other words, the volume of the divisor and of the total geometry are correlated. On the other hand, a divisor that is rigid can be shrunk to zero size without affecting the overall volume. Thus one would expect less of a correlation between a rigid divisor and the overall volume of the manifold.

Edge and face divisors are geometrically rigid (i.e. can have $\text{dim} \, H^1(D, \mathcal{O}_D)>1$) \cite{Braun:2017nhi}, while vertex divisors can be rigid or non-rigid. Therefore, it is natural to expect that vertex, edge, and face divisors will have qualitatively different behavior. Furthermore, just as the points in the polytope $\Delta^\circ$ correspond to divisors in $X$, edges in the triangulation $\mathcal{T}$ correspond to curves in $X$. In particular, if two points $\vec{p}_i$ and $\vec{p}_j$ are connected by an edge in a two-face in $\mathcal{T}$, then the corresponding divisors $D_i$ and $D_j$ intersect along a non-trivial curve in $X$. Because the intersection structure of vertex, edge, and face divisors is inherently different, it is reasonable to expect that their volumes will exhibit distinct behavior. Indeed, in \S\ref{sec:universal}, we will see that the full distribution of divisor volumes is made up of three sub-distributions, corresponding to each of the vertex, edge, and face divisor volumes.

\subsection{Calabi-Yau geometry} \label{subsec:cygeometry}

Having described the construction of Calabi-Yau threefolds as hypersurfaces in toric varieties, we will now describe some aspects of the resulting geometry that will be useful in deriving axion EFTs in \S\ref{subsec:axEFTs}.

A Calabi-Yau threefold $X$ is endowed with a K\"ahler form, $J$, which determines the metric on moduli space, as well as the volumes of minimum-volume representatives of all calibrated even-dimensional cycles in $X$:
\begin{align}
    \mathrm{vol}(C) = \int_{C} J, \ \ \ 
    \mathrm{vol}(D) = \frac{1}{2} \int_{D} J \wedge J, \ \ \ 
    \mathrm{vol}(X) = \frac{1}{6} \int_{X} J \wedge J \wedge J,
    \label{eq:volsJ}
\end{align}
where $C$ is an effective curve (2-cycle) and $D$ is an effective divisor (4-cycle).

The curves and divisors are characterized by integer sites in the $H_2(X,\mathbb{Z})$ and $H^2(X,\mathbb{Z})$ lattices, respectively, with $\text{dim}(H_2(X,\mathbb{Z})) =: h^{1,1}$. Not every site in the lattice is populated by a calibrated cycle: the cone of calibrated curves is the Mori cone, $\mathcal{M}_X$, and the cone of calibrated divisors is the cone of effective divisors, $\mathcal{E}_X$.\footnote{Not every site in $\mathcal{M}_X$ is calibrated, nor is every site in $\mathcal{E}_X$ calibrated, either (see \cite{holes} for a detailed study).} This work is concerned with the volumes of calibrated divisors\footnote{Non-calibrated divisors may also contribute to axion effective theories, but it is expected that their contributions to the four-dimensional effective theory are suppressed absent large recombination effects \cite{Demirtas:2021gsq}.}. In our study of Calabi-Yau threefolds constructed as hypersurfaces in toric varieties, we will focus on the subset of $\mathcal{E}_X$ that is generated by the set of prime toric divisors (see \S\ref{subsec:batyrev}), which are inherited from the ambient variety, and generate the cone $\mathcal{E}_V$ as a real cone. Note the relation $\mathcal{E}_V \subseteq \mathcal{E}_X$, as there can exist so-called \textit{autochthonous} divisors, which are not inherited from divisors in the ambient variety, and can live strictly outside of $\mathcal{E}_V$ \cite{Demirtas:2018akl}. We will restrict our analysis to Calabi-Yau threefolds defined via triangulations of favorable 4-dimensional reflexive polytopes \cite{batyrev1993dualpolyhedramirrorsymmetry} (see \S\ref{subsec:batyrev}), for which the number of prime toric divisors is $h^{1,1}+4$. 

The K\"ahler moduli space of $X$ is also a cone, known as the K\"ahler cone $\mathcal{K}_X$. $\mathcal{K}_X$ is the set of K\"ahler forms $J$ for which all even-dimensional cycles have non-negative volumes:
\begin{align}
    \mathcal{K}_X = \left\{J \left| \int_{C} J > 0 \right.\right\}.
\end{align}
On the interior of $\mathcal{K}_X$, all calibrated curves have strictly positive volumes, while on the boundary of $\mathcal{K}_X$, some curves shrink to zero size. The \textit{extended} K\"ahler cone, $\mathcal{K}$, is the union of the K\"ahler cones of all Calabi-Yau threefolds in the same birational equivalence class as $X$. On the interior of $\mathcal{K}$, all calibrated four-cycles have strictly positive volumes, while on the boundaries there exist four-cycles with vanishing volumes.

To evaluate volumes of submanifolds in $X$, one must specify a K\"ahler form $J$, i.e. a point in $\mathcal{K}_X$. $\mathcal{K}_X$ can be parameterized by a set of K\"ahler parameters, $t_i$, with $i = 1, \ldots h^{1,1}$. The K\"ahler form $J$ can be expanded in terms of these parameters:
\begin{align}
    J = t_i [D^i],
\end{align}
where $[D^i]$ is a basis for $H^2(X,\mathbb{Z})$. Plugging this into \eqref{eq:volsJ}, we arrive at the following expressions for the volumes:
\begin{align}
    \label{eq:volumes}
    \text{vol}(C^a) = M^{aj} t_j, \ \ \ \tau^a := \text{vol}(D^a) = \frac{1}{2} \kappa^{ajk} t_j t_k , \ \ \ 
    \mathcal{V} := \text{vol}(X)= \frac{1}{6} \kappa^{ijk} t_i t_j t_k
\end{align}
where $M^{aj}$ is a matrix counting the generic number of intersections between $C^a$ and $D^j$, and $\kappa^{ijk}$ are the triple intersection numbers of $X$. Here, $a=1,\ldots, h^{1,1}+4$.

The main geometric quantities of interest for this work are the prime toric divisor volumes $\tau^a$ of a given Calabi-Yau threefold. For use in \S\ref{sec:axiondivs}, \S\ref{sec:universal}, and \S\ref{sec:model}, we define the ``normalized divisor volumes"
\begin{align} 
    \hat{\tau}^a := \frac{\log(\tau^a)}{\langle \log(\vec{\tau})\rangle_X}
    \label{eq:normdivs}
\end{align}
where the average in the denominator is taken over all prime toric divisors in $X$.

Another geometric quantity of interest is the metric on moduli space, the K\"ahler metric $K_{ij}$. This metric is given by 
\begin{align}
    \label{eq:K_ij}
    K_{ij} :=  \frac{t_i t_j}{8\mathcal{V}^2} - \frac{A_{ij}}{4\mathcal{V}}
\end{align}
where $A_{ij} = \frac{\partial t_j}{\partial \tau^i}$. This metric will turn out to govern the kinetic terms of the axions, as we will explain in \S\ref{subsec:axEFTs}.

Finally we note that both the divisor volumes $\tau^a$ and the K\"ahler metric $K_{ij}$ scale homogenously along a given ray in $\mathcal{K}_X$. In other words, under the action
\begin{align}
    t_i \rightarrow \lambda t_i,
\end{align}
we have that
\begin{align}
    \tau^a &\rightarrow \lambda^2 \tau^a,\\
    K_{ij} &\rightarrow \lambda^{-4} K_{ij}.
\end{align}
We therefore note that once we have specified a spectrum of divisor volumes or a metric at some point $\vec{t} \in \mathcal{K}_X$, it is trivial to determine the volumes and K\"ahler metric along the entire ray $\lambda \vec{t}$.

In the following section, we will explain how the geometric data laid out here enter into the axion effective theories resulting from compactifications of string theory on toric hypersurface Calabi-Yau threefolds.

\subsection{Axion effective theories} \label{subsec:axEFTs}
In this section, we review the axion effective theories that arise from compactifications of string theory, as well as the geometric data that governs their kinetic terms and potentials.

Consider type IIB string theory compactified on a Calabi-Yau threefold (orientifold), $X$. The four-dimensional low energy effective theory is characterized by the fluxes in the internal manifold and the geometry of the manifold. 

Without the introduction of an orientifold, the resulting theory enjoys $\mathcal{N}=2$ supersymmetry, meaning that it has $8$ supercharges. This theory has massless fields corresponding to the $h^{2,1}$ complex structure moduli and $h^{1,1}$ K\"ahler moduli, where $h^{2,1}$ and $h^{1,1}$ are the Hodge numbers of $X$. Constraints on fifth forces dictate that for such a compactification to be compatible with our universe, these moduli fields must obtain a mass. To this end, an orientifold of $X$ is introduced with flux, breaking the supersymmetry to $\mathcal{N}=1$ and generating a potential for the moduli. In general, this orientifold projects out some number of moduli, leaving $h^{1,1}_+$ K\"ahler moduli. We will make the simplifying assumption that the orientifold projects out no K\"ahler moduli, so that $h^{1,1}$ complexified K\"ahler moduli fields appear in the effective theory (see \cite{Moritz:2023jdb, Sheridan:2024vtt} for work on Calabi-Yau orientifolds and their implications for axion physics).

Furthermore, in this work, we assume a moduli-stabilization scheme in which the complex structure moduli are stabilized at some high scale by flux and the real parts of the K\"ahler moduli (the saxions) are stabilized non-supersymmetrically at a somewhat  lower scale, leaving a low-energy effective theory for the imaginary parts of the K\"ahler moduli (the axions) with masses and couplings specified by the vacuum expectation values of the saxions. These vacuum expectation values correspond to the volumes of four-cycles (divisors) in $X$, and therefore an understanding of the axion effective theories requires an understanding of the geometry of $X$. Our aim here is to review how the axion EFT is derived from geometric quantities that we can compute.

In this setting, the K\"ahler moduli take the form:
\begin{align}
    T^j := \tau^j + i\theta^j := \frac{1}{2} \int_{\Sigma^j} J \wedge J + i \int_{\Sigma^j}C_4
\end{align}
where $J$ is the K\"ahler form on $X$, $C_4$ is the Ramond-Ramond four-form field, and the $\Sigma^i$ are an $h^{1,1}$-dimensional basis of four-cycles in $X$.  This theory enjoys $\mathcal{N}=1$ supersymmetry and is characterized in terms of a K\"ahler potential and superpotential (we work in units where $M_{\text{pl}}=1$):
\begin{align}
    K = -2\log (\mathcal{V})
    \label{eq:kahlerpotential}
\end{align}
and
\begin{align}
    W = W_0 + \sum_p A_p e^{-2\pi Q^p_{i} T^i},
    \label{eq:superpotential}
\end{align}
where the $A_p$ are the one-loop Pfaffians, which generically depend on the complex-structure moduli, and the $Q^p_i$ are the charges (taking values in $H^2(X,\mathbb{Z})$) of the instantons that contribute to $W$. Because this work is concerned with the effect of geometric structures on the axion low energy effective theory, we will set $A_p = W_0 = 1$ for simplicity. See \cite{Sheridan:2024vtt} for a more careful treatment of the effect of varying the magnitude of $W_0$ on the axion EFT. 

The resulting four-dimensional effective scalar potential is given by
\begin{align}
    V_F := e^K \left( K^{a\bar{b}} D_a W D_{\bar{b}} \overline{W} -3 |W|^2 \right)
    \label{eq:fterm}
\end{align}
where $D_a := \partial_a + \partial_a K$ is the covariant derivative on the field space.

The real parts of the K\"ahler moduli are the so-called saxions, and are scalar fields that are massless at the $\mathcal{N}=2$ level. As previously explained, in order to evade fifth force bounds, these saxions must be given a mass via some moduli stabilization scheme. In the context of this work, we will assume that the saxions are stabilized perturbatively, and that their vacuum expectation values comprise a dense set in $\mathcal{K}_X$. We will therefore implicitly assume that for any choice of $\tau^i$, there exists an effective potential given by \eqref{eq:fterm} in which the saxions are stabilized at their vevs. The imaginary parts of the $T^i$ are axions and are the dynamical fields whose properties we are interested in in this work. It is important to emphasize that different moduli stabilization schemes will lead to qualitatively different distributions for cycle volumes. For example, in the KKLT scenario \cite{kklt}, the moduli must be fixed at a place in $\mathcal{K}_X$ where most of the divisor volumes are approximately the same size, so that they may all compete with the flux-induced superpotential. We focus here, rather, on ``generic" locations in moduli space. In \S\ref{sec:axiondivs} we will explain precisely how we sample these points.

The effective theory for these axions takes the form 
\begin{align}
    \label{eq:axioneft}
    \mathcal{L} = -\frac{1}{2} K_{ij} \partial_\mu \theta^i \partial^\mu \theta^j -V(\vec{\theta}) + \text{couplings to Standard Model}
\end{align}
where the potential is given by \eqref{eq:fterm} and has the structure
\begin{align}
    \label{eq:axionEFT_potential}
    V(\theta^i) = \sum_I \Lambda_I^4 \left[1 - \cos(2\pi Q^I_{i} \theta^i + \delta_I)\right].
\end{align}
Here, the $\Lambda_I^4$ are exponentially sensitive to the values of the $\tau^i$ as in \eqref{eq:superpotential} and the $\delta_I$ are phases determined by the phases of $A_p$ and $W_0$. 

Using \eqref{eq:kahlerpotential} and \eqref{eq:superpotential}, we approximate the coefficients in \eqref{eq:axionEFT_potential} as
\begin{align}
    \Lambda_{I}^4 \approx 8 \pi e^{K/2} \frac{\vec{q}_I \cdot \vec{\tau}}{\mathcal{V}} e^{-2\pi \vec{q}_I \cdot \vec{\tau}}
\end{align}
where now the index $I$ runs over the prime toric divisors in $X$. Note that here we focus on single-instanton terms as these are the leading terms. We now explain how to compute axion masses and decay constants, starting from the axion effective theory \eqref{eq:axioneft}.

Axion decay constants have been defined in the literature in a number of often incompatible ways. In this work, we will define the decay constant $f_i$ of an axion $\phi_i$ as the field range of $\phi_i$ in the basis where $\phi_i$ is an approximate mass eigenstate. Then to compute the axion decay constants (and subsequently their masses), we rotate the basis to one in which the axions are simultaneous kinetic and mass eigenstates, following the algorithm laid out in \cite{Gendler:2023kjt}. The decay constants are then
\begin{align}
    f_i = \frac{1}{2\pi} \left[Q_{ij} (M^{-1})^j_{\, i}\right]^{-1},
    \label{eq:decayconst}
\end{align}
where $M_i^j$ is the matrix that canonically normalizes the axion kinetic term, i.e. $\phi_i = M_i^j \theta_j$ with $\phi_i$ the canonically normalized axions.

As a distribution, the $f_i$ in \eqref{eq:decayconst} are readily approximated by the eigenvalues of the K\"ahler metric \eqref{eq:K_ij}:
\begin{align}
    \{f_i\} \approx \sqrt{\text{eig}(K_{ij})}.
    \label{eq:f_K}
\end{align}
In fact, often \eqref{eq:f_K} is taken as a definition of the axion decay constants. In \S\ref{sec:axiondivs}, we will find useful approximations for $\text{eig}(K_{ij})$ and see that we are able to generally reproduce the distributions $\{f_i\}$.

The axion masses are then:
\begin{align}
    m_i^2 = \frac{\Lambda_i^4}{f_i^2}.
    \label{eq:mass}
\end{align}

In \S\ref{sec:model}, we will aim to reproduce scalings of axion observables as a function of $h^{1,1}$ obtained in previous literature \cite{Demirtas:2018akl, Mehta:2021pwf, Halverson:2019cmy, Gendler:2023kjt, Demirtas:2021gsq, Gendler:2024adn} using a simplified model. In addition to reproducing the masses and decay constants, as described above, we will also aim to reproduce the observed scaling of the stringy contribution to the QCD $\theta$-angle found in \cite{Demirtas:2021gsq} with our model. We will thus now describe how to compute this quantity in the context of type IIB compactifications (see \cite{Demirtas:2021gsq} for more detail).

To start, we suppose that for each Calabi-Yau threefold $X$, there exists a divisor $D_{\text{QCD}}$ on which a toy model of QCD is hosted via a stack of D7-branes. The QCD axion is identified with the Ramond-Ramond four-form potential, integrated over $D_{\text{QCD}}$:
\begin{align}
    \theta_{\text{QCD}} = \int_{D_{\text{QCD}}} C_4.
\end{align}

If this axion solves the strong CP problem, then the contribution to its potential from low-scale QCD instantons must dominate over the stringy instantons, coming from Euclidean D3-branes wrapped on divisors. In particular, to conclude that $\theta_{\text{QCD}}$ solves the strong CP problen, we require
\begin{gather}
    \label{eq:Delta_theta}
\Delta\theta_{\text{stringy}}\approx \frac{\Lambda_{\vec{q}_c}^4}{\Lambda_{\text{QCD}}^4}<10^{-10},
\end{gather}
where $\Lambda_{\vec{q}_c}^4$ is the instanton scale associated with the smallest-action instanton such that $\vec{q}_c$ and all more-dominant instantons contains the charge associated with the QCD axion, $\vec{q}_{\text{QCD}}$ in their linear span. In \cite{Demirtas:2021gsq}, it was shown that $\Delta \theta_{\text{stringy}}$ decreases rapidly with $N$, the number of axions. We will aim to reproduce this behavior in \S\ref{sec:model}.

\section{Axion properties from divisor volumes} \label{sec:axiondivs}

In this section, we will argue that the divisor volumes are sufficient to approximately reproduce the data characterizing the axion effective theory \eqref{eq:axioneft}. To do this, we will first construct an ensemble of Calabi-Yau threefolds as hypersurfaces in toric varieties, explaining how we attempt to make this ensemble a ``fair" sample of the set of such manifolds. We will then explain the approximations we make to the axion observables and compare these approximations to the exact observables computed using \eqref{eq:axioneft}.

\subsection{An ensemble of Calabi-Yau threefolds} \label{subsec:ensemble}

The goal of this work is to identify universal features in the set of Calabi-Yau threefolds constructed as toric hypersurfaces that shed light on patterns observed in the resulting axion physics of string compactifications on these manifolds. To do this, one wants to study the relevant geometry in a representative sample of toric hypersurface Calabi-Yau threefolds. This task is obscured by the fact that the set of such manifolds is enormous: there are 473,800,776 in the Kreuzer-Skarke database, and each one can be triangulated a number of ways to define a toric variety in which the anti-canonical hypersurface gives rise to a Calabi-Yau threefold, as explained in \S\ref{subsec:batyrev}. The number of distinct Calabi-Yau threefolds obtainable from this database is not known, but an upper bound is given by $10^{428}$ \cite{Demirtas:2020dbm}. The size of this set makes a comprehensive analysis intractable, so in this work (and in previous analyses of this database \cite{Demirtas:2018akl, Gendler:2023kjt, Gendler:2024adn, Mehta:2021pwf, Gendler:2023hwg, Demirtas:2021gsq}) we will rely on sampling this set as systematically as possible.

When undertaking the task of studying statistics of observable properties in the string landscape, one must grapple with the problem of how to weight the various vacua in a way that yields useful results about the physics. Even if one were to assume that the set of vacua was finite and tractably small, the various measures that one can put on the space yield different results for physical predictions. In a world with infinite computational power, one would compute the full set of four-dimensional string vacua, with all of their associated dynamical transitions and tunneling rates, and use this data to assign a cosmological measure on the space (see e.g. \cite{Bousso:2008hz}). In reality, constructing string vacua is only possible in a number of limited ways on a case by case basis (see e.g. \cite{Demirtas:2021nlu, Rajaguru:2024emw}), and calculating a cosmological measure is far beyond current capabilities. We will now explain the approach we take instead.

Instead of constructing the complete set of string theory vacua, we will consider the set of solutions that arise from compactifying type IIB string theory on a Calabi-Yau threefold constructed as toric hypersurfaces. We will assume that the true vacua are densely populated within the moduli spaces of these manifolds, and therefore take a random sample with a flat measure in this space as representative. For our purposes, a given ``vacuum" consists of a specification of three objects: 
\begin{itemize}
    \item a  polytope $\Delta^\circ$ from the Kreuzer-Skarke database,
    \item a  triangulation $\mathcal{T}$ of $\Delta^\circ$, defining a Calabi-Yau threefold $X$, and
    \item a point $\vec{t}$ in the K\"ahler moduli space of $X$, $\mathcal{K}_X$.
\end{itemize}
Our objective in looking for universal features across this landscape is to produce as fair a sample as possible from within each of these categories. We will now explain our methods for each.

Obtaining a fair sample of the set of polytopes is straightforward: the number of four dimensional reflexive polytopes is finite and small enough that a random subset is possible to obtain. To this end, we analyzed polytopes with
\begin{gather}
    h^{1,1} = \{20, 30, 40, 50, 60, 80, 100, 120, 150, 180, 200\}.
\end{gather}
For $h^{1,1} \leq 120$, we obtained a sample of $10^4$ randomly selected polytopes per $h^{1,1}$. For polytopes with $h^{1,1}>120$ we included every polytope in our ensemble.

Constructing a fair sample of Calabi-Yau threefolds from triangulations of the set of polytopes is much more elusive. The putative number of FRSTs of a given polytope  grows exponentially with $h^{1,1}$ \cite{Demirtas:2020dbm}, and construting random subsets of the FRSTs by first constructing \textit{all} FRSTs quickly becomes impossible. Furthermore, even if an exhaustive list of FRSTs was possible to obtain, the set of unique Calabi-Yau threefolds defined by this set could be quite a lot smaller. For instance, as explained in \cite{Demirtas:2020dbm}, two FRSTs with the same set of two-face triangulations give rise to homotopy-equivalent Calabi-Yau threefolds. Furthermore, even FRSTs with different two-face triangulations can give rise to equivalent manifolds in some cases, as explored in \cite{gendler2023counting, Chandra:2023afu}. For modest-to-large $h^{1,1}$, finding the set of topologically inequivalent Calabi-Yau threefolds from a set of FRSTs is computationally intractible. This poses an obstacle in sampling, because if many FRSTs correspond to small numbers of distinct Calabi-Yau threefolds, sampling from the space of FRSTs likely oversamples instances of certain manifolds. Furthermore, at large $h^{1,1}$ there is a danger with any sampling method that the sampled distribution does not match the true distribution, since the underlying distribution is unobtainable. Finally, any sampling method that relies in some way on sampling points in the extended K\"ahler cone is in danger of producing a biased sample, for the following reason: nearby K\"ahler cone chambers within $\mathcal{K}$ are more likely to be ``similar" geometries, in that they differ only by a small number of effective curves. To mitigate the effects of the potential for sampling bias in sampling multiple triangulations of a given polytope, we will use the sampling method described below to sample a single triangulation of each polytope, relying on the diversity of favorable, 4D reflexive polytopes in the KS database to furnish our ensemble.

Sampling methods broadly fall into two categories. The first relies on exploring the space of triangulations via direct constructions and manipulations of triangulations of polytopes as a collection of points and edges. The second method involves exploring the extended K\"ahler cone of a given Calabi-Yau via continuous deformations of the moduli. Prototypes of these methods are outlined in \cite{Demirtas:2020dbm} as \texttt{Algorithm 1} and \texttt{Algorithm 2}\footnote{See \cite{Yip:2025hon} for a recent approach.}.

Our ensemble is constructed using a version of \texttt{Algorithm 2}. The basic idea of the algorithm is to perform a random walk within the extended K\"ahler cone of a given Calabi-Yau. We implement this algorithm, modified so that the random walk is performed along straight line coordinates parameterized by divisor volumes\footnote{We thank Jakob Moritz for providing the code for this method.}. For each polytope, we select a single Calabi-Yau threefold, randomly selected in this way.

Finally, we turn to the sampling of points $\vec{t}$ inside $\mathcal{K}_X$. Even within the moduli space of a particular Calabi-Yau threefold, there are different measures that one can put on the space for sampling. A natural measure one can use is the Weil-Petersson measure, which is given by the determinant of the K\"ahler metric\footnote{Sampling with this measure will be explored in \cite{pipeline}.}.  As explained in \S\ref{subsec:cygeometry}, divisor volumes are homogenous under rescalings of the K\"ahler parameters. In other words, specifying a set of divisor volumes at a point $\vec{t}_0$ specifies the spectrum of divisor volumes at every point along the ray $\lambda \vec{t}_0$. Therefore, our sample of points in K\"ahler moduli space will be restricted to an angular slice of $\mathcal{K}_X$. We will sample these points by using the random walk method described above. In other words, we perform a random walk in divisor coordinates in the extended K\"ahler cone. Endpoints of this random walk specify both a Calabi-Yau threefold and a point in $\mathcal{K}_X$.

Our final dataset is obtained by randomly selecting $O(10^4)$ polytopes per $h^{1,1}$ up to $h^{1,1}=120$, and using all the available favorable polytopes from $h^{1,1}=150$ to $h^{1,1}=200$ (which is less than $10^4$ per $h^{1,1}$). For each polytope, we pick one random triangulation with \texttt{Algorithm 2} and one random point in the K\"ahler moduli space of the corresponding Calabi-Yau threefold.

A remark is in order concerning the susceptibility of our ensemble to the sampling biases described above. The goal of this work is to use a consistent method to construct ensembles of Calabi-Yau geometries in order to understand what, if any, the universal features in these datasets are. More work is warranted to finalize a systematic sampling method that is guaranteed to generate fairly sampled distributions. Here, we endeavor to generate such ensembles to the extent that we can with finite computation time and investigate what lessons can be learned from these datasets, with the caveat in mind that distributions may change as the sampling method is changed. However, we do not expect broad qualitative results to be sensitive to these inherent biases. The first reason for this is that although different sampling methods produce slightly different distributions for divisor volumes, the general behavior (roughly speaking, that hierarchies in four-cycle volumes increases as a function of $h^{1,1}$) persists independent of the method. Furthermore, recently machine learning techniques have been employed \cite{MacFadden:2024him} with the goal of finding large deviations in axion properties from those that have been previously studied, and have been unable to find large deviations from previous results, even with robust optimization schemes. In \cite{Sheridan:2024vtt}, it was shown analytically that large deviations in axion observables \textit{can} be found, but that such regions of the moduli space are rare (with respect to the Weil-Petersson metric) and cannot be recovered with simple sampling or machine learning techniques. Nevertheless, we caution that our dataset may not be a fair sample of Calabi-Yau threefolds and that a part of the space of all manifolds may be sampled more often than the others.

To corroborate our choice of sampling method, we performed a systematic comparison of an ensemble of Calabi-Yau threefolds constructed as described above with a comprehensive set of geometries at sufficiently small $h^{1,1}$. In particular, every FRST for every polytope with $h^{1,1} \leq 7$ in the Kreuzer-Skarke database was obtained in \cite{Sheridan:2024vtt}. We compared our ensemble to a random sample of this exhaustive list and found that the distributions of divisor volumes, overall volumes, and K\"ahler metric eigenvalues match.

\subsection{Statistical tests} \label{subsec:stats}
 A main concern of this work is to diagnose the similarity of divisor volume distributions across different Calabi-Yau threefolds (see \S\ref{sec:universal}), as well as to ultimately compare distributions of axion observables obtained using a simplified model with the true values in our ensemble of compactifications (see \S\ref{sec:model}). To this end, we perform statistical tests that that compare two samples of data and determine the probability that they are drawn from a unified distribution. Let us briefly describe the tests we use, and then discuss the results.

As a measure of distance between two probability distributions, throughout this work we implement a computation of distance using the Wasserstein metric \cite{monge1781memoire,kantorovich1942translocation}, specifically computing the $1$-Wasserstein distance, also known as the Earth Mover's distance. Distances with respect to this metric intuitively measure how much work it takes to transform one probability distribution into another, taking into account both the difference in shape of the distribution as well as the spatial separation. Given two probability distributions defined over $x\in\mathbb{R}$, $p_1(x)$ and $p_2(x)$, whose corresponding cumulative distribution functions are $F_1(x)$ and $F_2(x)$, the $1$-Wasserstein distance is defined as \cite{vallender1974calculation}
\begin{gather}
    \label{eq:W1}
    W_1(p_1,p_2):=\int_{-\infty}^\infty dx|F_1(x)-F_2(x)|.
\end{gather}
Note that probability distributions can be empirically obtained from discrete datasets, as in our case. For two sorted discrete datasets $X=\{X_i\}$ with $X_i\leq X_2\leq\cdots\leq X_n$ and $Y=\{Y_i\}$ with $Y_1\leq Y_2\leq\cdots\leq Y_m$, their empirical CDFs are computed by
\begin{gather}
    F_X(z)=\frac{\text{number of }X_i\leq z}{n}\quad\quad
    F_Y(z)=\frac{\text{number of }Y_j\leq z}{m}.
\end{gather}
Combining this with \eqref{eq:W1}, we get an explicit expression for the $1$-Wasserstein distance between $X$ and $Y$:
\begin{gather}
    W_1(X,Y)=\sum_{k=1}^{K-1}(Z_{k+1}-Z_k)|F_X(Z_k)-F_Y(Z_k)|,
\end{gather}
where
\begin{gather}
    Z=\{Z_1,Z_2,\cdots,Z_K\}=\text{sort}\{X_1,X_2,\cdots,X_n,Y_1,Y_2,\cdots,Y_m\}.
\end{gather}
In order to gauge the performance of fit functions over different variables, we present our results in terms of a relative percentage error defined as
\begin{gather}
   \epsilon_W := \frac{W_1(X,Y)}{\text{mean}(\overline{X},\overline{Y})}\times100
\end{gather}
for two datasets $X$ and $Y$, where $\overline{X}$ and $\overline{Y}$ are their mean values.

Another statistical test that will be useful when we compare different fit models is the adjusted $r^2$ test. This test is similar to the commonly used $r^2$ goodness-of-fit test, but has the advantage that it prevents overfitting by penalizing fit functions with many parameters. For a dataset of size $n$ with original values $\{y_1,y_2,\cdots,y_n\}$ and values obtained using a fit function $\{y_1^\text{fit},y_2^\text{fit},\cdots,y_n^\text{fit}\}$, $r^2$ is defined as
\begin{gather}
    r^2:= 1-\frac{\sum_{i=1}^n(y_i^\text{fit}-y_i)^2}{\sum_{i=1}^n(y_i-\bar{y})^2},
\end{gather}
where
\begin{gather}
    \bar{y}:= \frac{1}{n}\sum_{i=1}^n y_i
\end{gather}
is the mean value. The best fitting case results in  $r^2=1$ while a bad fit may give $r^2<0$ (intuitively, a negative $r^2$ value suggests that the value $\bar{y}$ alone gives a better fit to the data). The adjusted $r^2$ is defined as
\begin{gather}
    r^2_\text{adj}:= 1-(1-r^2) \left(\frac{n-1}{n-p-1}\right)
\end{gather}
with $p$ being the number of parameters in the fit function that excludes constant terms. By definition, $r^2_\text{adj}\leq r^2$, and the two values differ more greatly when $p$ is large.

\subsection{Comparison of exact observables with divisor-volume approximations} \label{subsec:divapprox}

In this section, we justify the approximations that we will make in constructing our simplified model in \S\ref{sec:model}. In particular, we argue that divisor volumes alone are enough to approximate the axion observables described in \S\ref{subsec:axEFTs}.

One of the main observables that we are interested in replicating with a simple model is the set of axion decay constants. \eqref{eq:decayconst} gives an expression for computing axion decay constants as approximate field ranges of the axion potential, but we will now argue that the distribution of decay constants can be approximated in a relatively simple manner in terms of divisor volumes.

To approximate the eigenvalues of $K_{ij}$, we first note that in \eqref{eq:K_ij}, the first term is suppressed by one power of the overall volume, as compared to the second term. Secondly, we observe in our ensemble that $A_{ij}$ is almost diagonal, so that the eigenvalues of the K\"ahler metric are well approximated by
\begin{align}
    \text{eig}(K_{ij}) \approx \frac{|A_{ii}|}{4\mathcal{V}}.
\end{align}
Finally, we would like to approximate this quantity in terms of divisor volumes. To this end, we note the relations
\begin{align}
    A_{ii} = \frac{t_i}{2 \tau^i} \ \ \ \text{and} \ \ \ 
    \mathcal{V} = \frac{1}{3} \tau^k t_k.
\end{align}
To estimate these quantities, we note that if the triple intersection matrix were diagonal, we would have
\begin{align}
    \tau^i = \frac{1}{2}\kappa_{iii} (t^i)^2, \ \ \ \mathcal{V} = \frac{1}{6} \sum_i \kappa_{iii} (t^i)^3
\end{align}
so that in particular we can write
\begin{align}
    t_i = \sqrt{\frac{2 \tau^i}{\kappa_{iii}}}.
\end{align}
To approximate $\mathcal{V}$, we note that in our ensemble the triple intersection numbers $\kappa_{iii}$ are $\mathcal{O}(1-10)$ and the $t_i$ exhibit large hierarchies: the largest $t_i$ grows with $h^{1,1}$, and can be as large as $\pm10^4$ for $h^{1,1}\sim200$.  We therefore motivate the approximation
\begin{align}
    \mathcal{V} \approx \frac{1}{\sqrt{8}} \kappa_{\text{max}}^{-2/3} \tau_{\text{max}}^{3/2}
\end{align}
where $\kappa_{\text{max}} = \kappa_{III}$ for $\tau_I = \tau_{\text{max}}$. In our ensemble this  becomes a better approximation as $h^{1,1}$ increases.

Putting these approximations together, and neglecting $\mathcal{O}(1)$ factors, we have 
\begin{align}
    \text{eig}(K_{ij}) \approx \frac{1}{\tau_{\text{max}}^{3/2} \sqrt{\tau_i} }
\end{align}
so that our final approximation for the set of axion decay constants is (using \eqref{eq:f_K})
\begin{align}
    f_i \approx \frac{1}{\tau_{\text{max}}^{3/4} \tau_i^{1/4}}.
    \label{eq:f_approx}
\end{align}

Note that in an explicit geometry, the above expressions involve choosing a basis of divisors $\tau_i$ from the over-complete set of divisors $\tau_I$. In our model in \S\ref{sec:model} based on the findings of \S\ref{sec:universal}, we will not obtain information about the charges of divisor volumes that we reconstruct. Therefore, one aspect of our approximation in this section is to neglect the dependence of axion observables on the charges of the divisors on which they depend. To this end, we make the further approximation
\begin{align}
    \tau_i =\tau_I
    \label{eq:basis_approx}
\end{align}
where we have ordered the volumes $\tau_I \leq \tau_{I+1}$ and the range of $I$ in \eqref{eq:basis_approx} is now $I = 1, \ldots, h^{1,1}$. In other words, we approximate the volumes of a basis of divisors as the volumes of the $h^{1,1}$ smallest prime toric divisors.

Next, we seek an approximation for the axion masses. We begin with the full potential given in \eqref{eq:axionEFT_potential}. As explained in \S\ref{sec:review}, we approximate the coefficients in \eqref{eq:axionEFT_potential} as 
\begin{gather}
    \Lambda_{I}^4 \approx 8 \pi \frac{\tau_I}{\mathcal{V}^2} e^{-2\pi \tau_I}
\end{gather}
where $I$ runs over each of the $h^{1,1}+4$ prime toric divisors. Using our approximation for the overall volume above (again, dropping $\mathcal{O}(1)$ factors), we have 
\begin{gather}
    \Lambda_{I}^4 \approx  \frac{\tau_I}{\tau_{\text{max}}^3} e^{-2\pi \tau_I}.
    \label{eq:l_approx}
\end{gather}
Plugging this expression, along with our approximation for the axion decay constants \eqref{eq:f_approx}, into the expression for the axion masses \eqref{eq:mass}, we obtain our final approximation:
\begin{align}
    m_i \approx \frac{\tau_i^{3/4}}{\tau_{\text{max}}^{3/4}} e^{-\pi \tau_i}.
    \label{eq:m_approx}
\end{align}

\begin{figure}[h!]
    \centering
    \includegraphics[width=\linewidth]{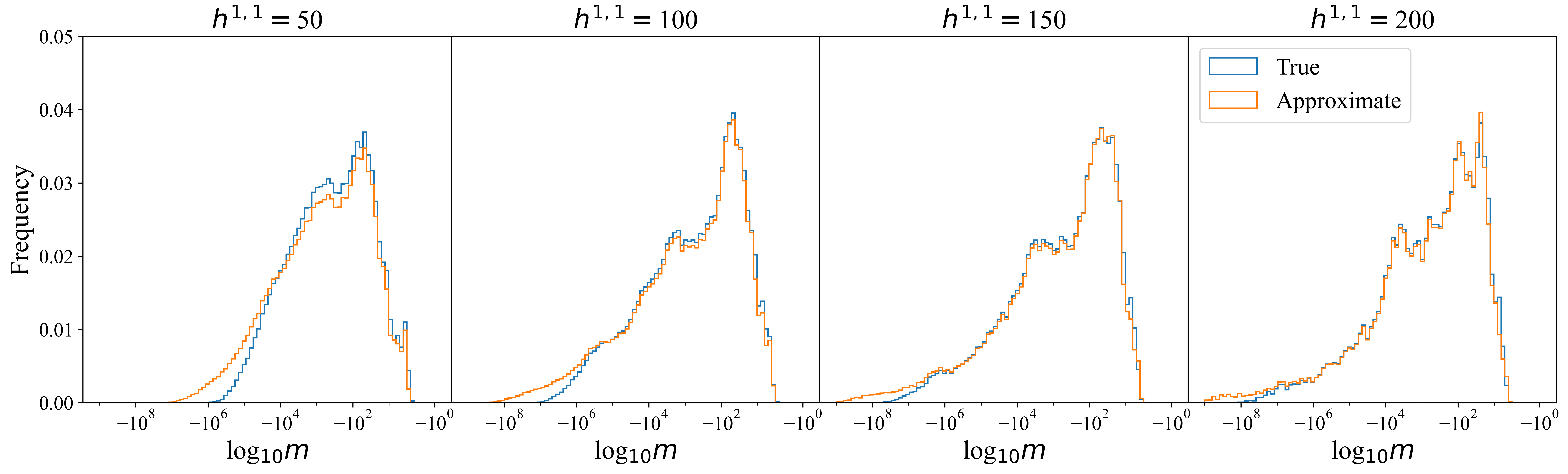}
    \includegraphics[width=\linewidth]{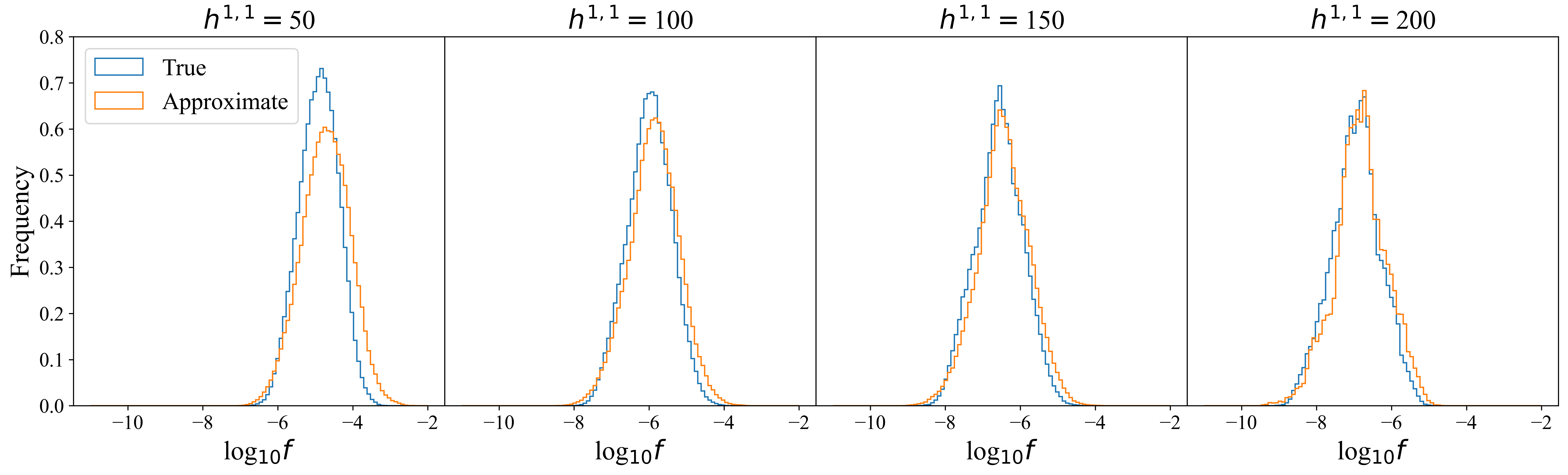}
    \caption{Top: Comparison of true and approximate distributions of axion masses for various values of $h^{1,1}$.
    Bottom: Comparison of true and approximate distributions of decay constants for various values of $h^{1,1}$. In the plots, ``true'' means \eqref{eq:mass} and \eqref{eq:decayconst} computed using the true eigenvalues of the K\"ahler metric, original volumes, and original charge matrices; ``approximate'' means using  \eqref{eq:f_approx} and \eqref{eq:m_approx}.}
    \label{fig:mf}
\end{figure}

\begin{figure}[h!]
    \centering
    \includegraphics[width=0.45\linewidth]{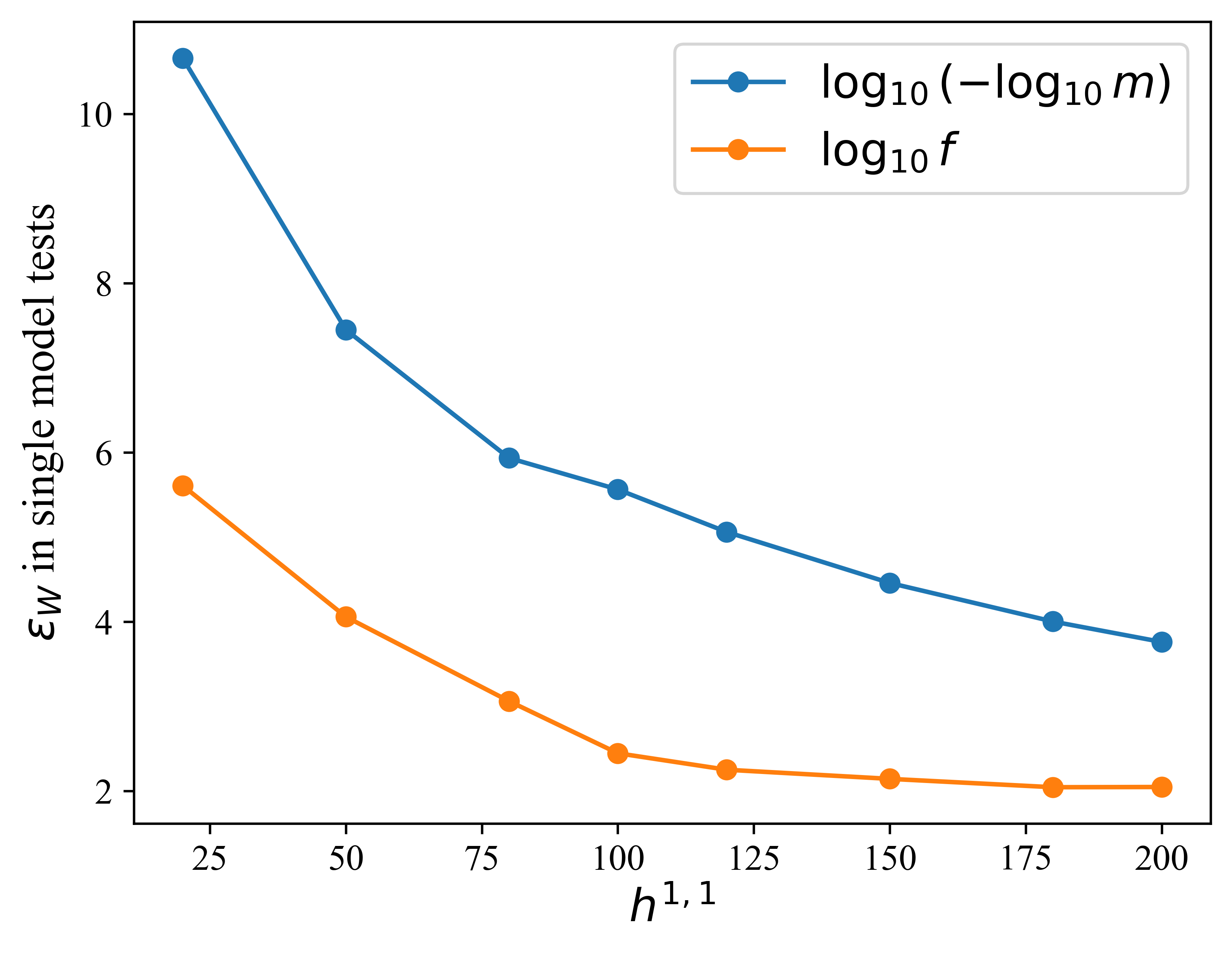}
    \includegraphics[width=0.45\linewidth]{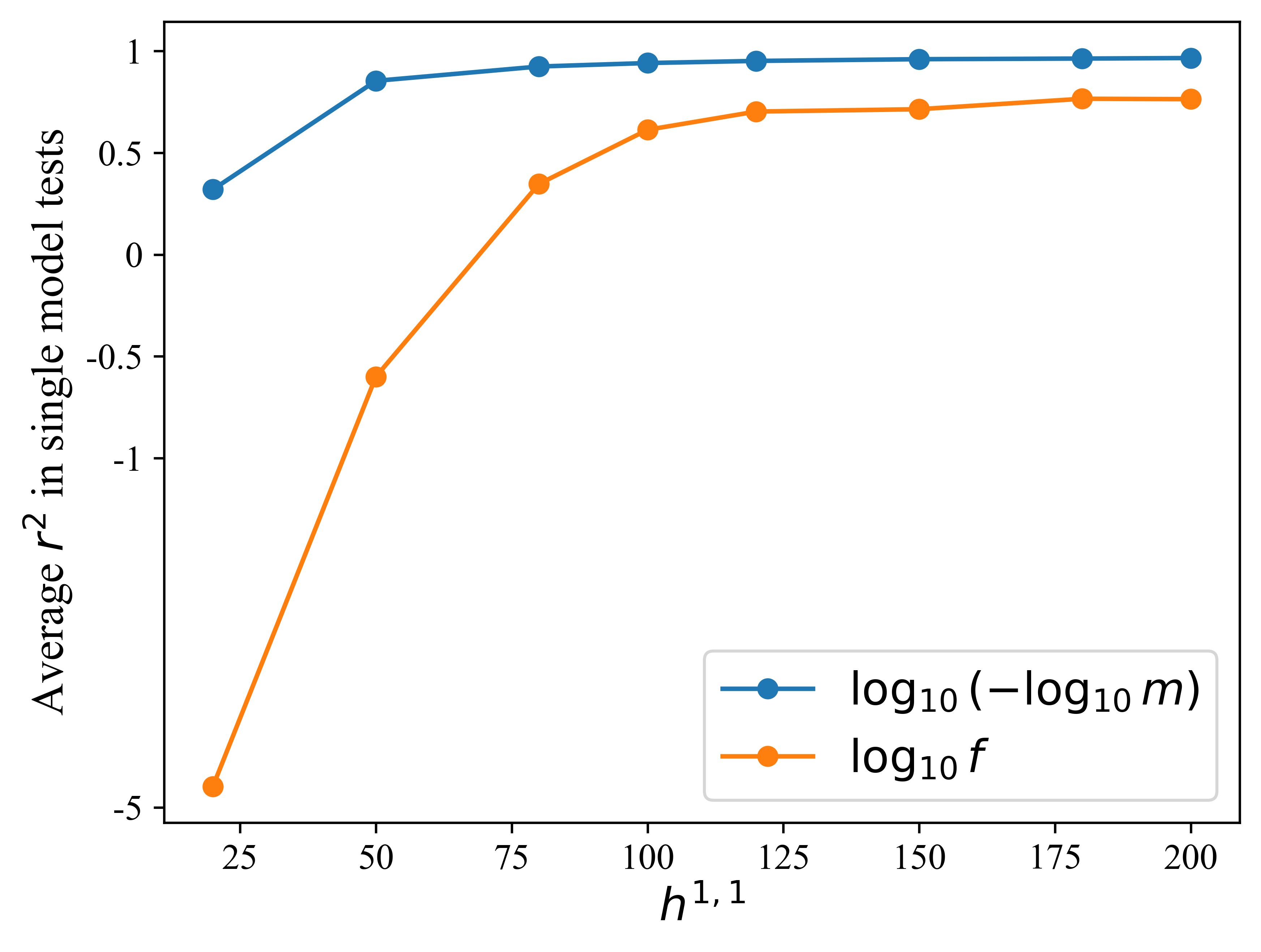}
    \caption{$\epsilon_W$ (left) and $r^2$ (right) between a true-valued model and its corresponding approximated model, averaged over 1000 single model tests, for axion masses and decay constants, respectively.
    }
    \label{fig:percentage_mf}
\end{figure}

Of course, the triple intersection numbers $\kappa_{ijk}$ are \textit{not} diagonal, and therefore \eqref{eq:f_approx} and \eqref{eq:m_approx} are just approximations. In fact, it is known that making the approximation that the triple intersection numbers are diagonal can lead to qualitatively different results for scaling laws than direct computation \cite{Long:2021jlv}. However, in this case we compare the estimates \eqref{eq:f_approx} and \eqref{eq:m_approx} to the actual values computed using \eqref{eq:decayconst} and \eqref{eq:mass} and the full set of geometrical data and find relatively good agreement. This comparison is shown qualitatively in Fig.~\ref{fig:mf} for several different values of $h^{1,1}$. Fig.~\ref{fig:percentage_mf} shows the percentage error in the average 1-Wasserstein distance between the true values and their approximations, as well as the $r^2$ values for the same comparison. Note that the percentage errors in the $1$-Wasserstein distance are larger for the mass distributions: this is because we are comparing these sets on a doubly log scale.  Based on these statistical tests, we conclude that \eqref{eq:f_approx} and \eqref{eq:m_approx} are adequate approximations of the true decay constants and masses for a particular compactification.

\begin{figure}[h!]
    \centering
    \includegraphics[width=\linewidth]{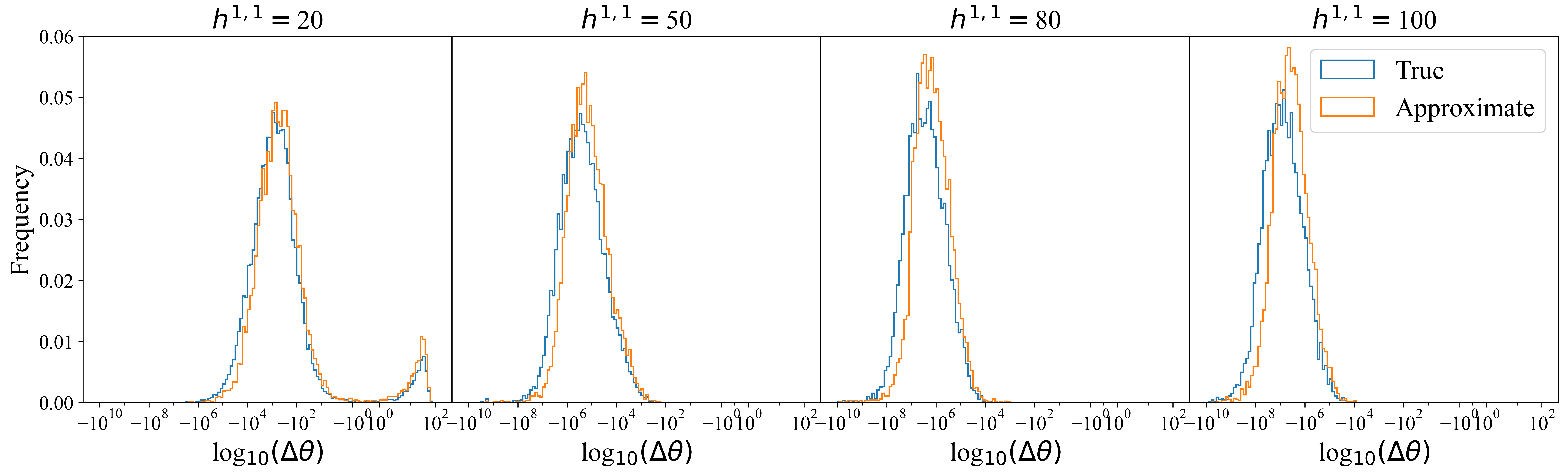}
    \caption{
    Comparisons of the true and approximate distributions of $\Delta\theta$. In the plots, “true” means using the original charge matrices as in \eqref{eq:Delta_theta}, and “approximate” means picking the $h^{1,1}$th largest $\Lambda_I$ as the dominant $\Lambda_I$ in \eqref{eq:Delta_theta}.
    }
    \label{fig:theta_compare}
\end{figure}

Another observable we will analyze in this work is the stringy contribution to the QCD $\theta$-angle, as discussed in \S\ref{subsec:axEFTs}. To this end, we use the approximation \eqref{eq:l_approx} to estimate \eqref{eq:Delta_theta}. An important aspect of this approximation is in choosing which instanton contribution to use as input in \eqref{eq:Delta_theta}. As explained in \S\ref{subsec:axEFTs}, the most dominant PQ-breaking instanton is obtained by finding the largest $\Lambda_I$ with associated charge $q_I$ such that $q_I$ and all charges associated to more dominant instanton scales contain the QCD charge in their span.  In our ensemble, we find that the $h^{1,1}$-largest instanton scale is most often the most dominant PQ-breaking scale (this is the case in well over $50\%$ of examples). We therefore find that an adequate simplification is to estimate the size of the dominant PQ-breaking instanton with the instanton associated to the $N^{\text{th}}$ smallest divisor volume, where $N=h^{1,1}$. This approximation is shown in Fig.~\ref{fig:theta_compare}, where it can clearly be seen that this simplification suffices to reproduce the distributions of $\Delta \theta$.

\section{Divisor spectra in toric hypersurfaces\label{sec:universal}}

In this section, we will provide evidence that the spectra of appropriately normalized divisor volumes in toric hypersurface Calabi-Yau threefolds share common underlying features. These features correspond to the distributions of the volumes of the vertex, edge, and two-face divisors, respectively.  We show that these distributions remain  roughly constant as one varies the point in moduli space, the Calabi-Yau threefold, and even the Hodge number $h^{1,1}$ of the threefold.

For each Calabi-Yau threefold, we are interested in the volumes of the prime toric divisors $\tau^I$ with $I=1,\dots,h^{1,1}+4$. The normalized and original divisor volumes are related by \eqref{eq:normdivs}:
\begin{align} 
    \label{eq:normed_tau}
    \hat{\tau}^I := \frac{\log(\tau^I)}{\langle \log(\vec{\tau})\rangle_X}
\end{align}
where 
\begin{gather}
    \label{eq:mean_log_tau}
   \langle \log(\vec{\tau})\rangle_X :=\frac{1}{h^{1,1}+4}\sum_{I=1}^{h^{1,1}+4}\log\left( \tau^I \right).
\end{gather}
Here, we have performed a rescaling such that 
\begin{align}
    \text{min}_I (\tau^I) = 1.
\end{align}
Recall that this rescaling is arbitrary because we can obtain the divisor volumes at any point along a ray in $\mathcal{K}_X$ just by an overall constant prefactor, as explained in \S\ref{subsec:cygeometry}.

Similarly, we can define normalized divisor volumes restricted to the vertex, edge, and two-face divisors:
\begin{align} 
    \label{eq:normed_tau_v}
    \hat{\tau}^a_{v,e,f} = \frac{\log(\tau^a_{v,e,f})}{\langle \log(\vec{\tau})\rangle_{v,e,f}}
\end{align}
where the subscripts indicate that the divisors are vertex, edge, and two-face divisors, respectively.

\begin{figure}[h!]
    \centering
    \includegraphics[width=0.6\linewidth]{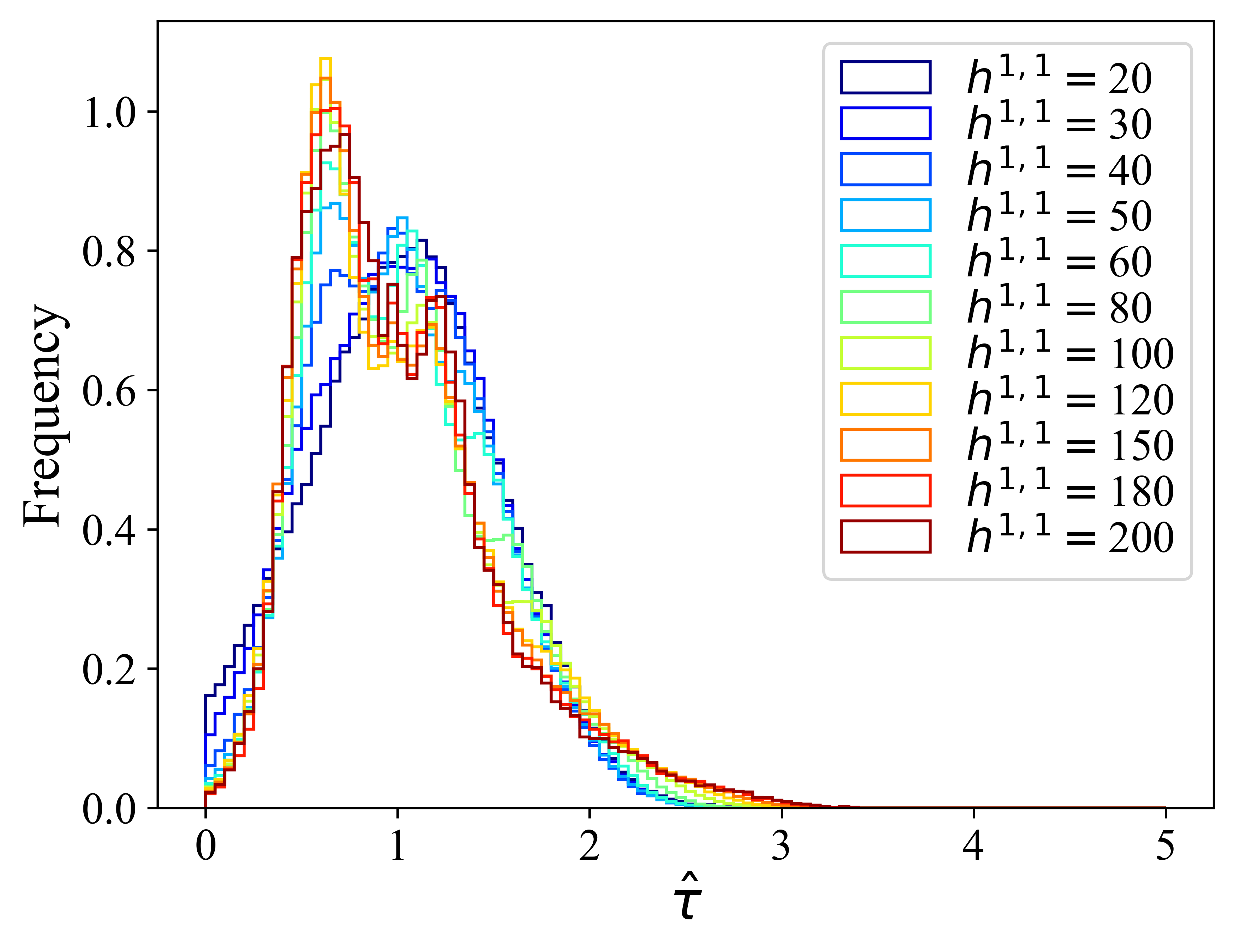}
    \caption{Distributions of all the normalized divisor volumes (excluding the minimum in each Calabi-Yau) of all the Calabi-Yau threefolds in our dataset, color coded by $h^{1,1}$.}
    \label{fig:tau_all}
\end{figure}

We then construct an ensemble of Calabi-Yau threefolds as described in \S\ref{subsec:ensemble}. For each geometry $X$, we compute the set of normalized divisor volumes $\{\hat{\tau}^I \}_X$ and $\{\hat{\tau}^a_{v,e,f} \}_X$. Figure~\ref{fig:tau_all} shows the distributions of $\hat{\tau}^I$ across the various values of $h^{1,1}$ in our ensemble. Each histogram represents the collection of all normalized divisor volumes for every Calabi-Yau with a fixed $h^{1,1}$.

\begin{figure}[h!]
    \centering
    \includegraphics[width=0.32\linewidth]{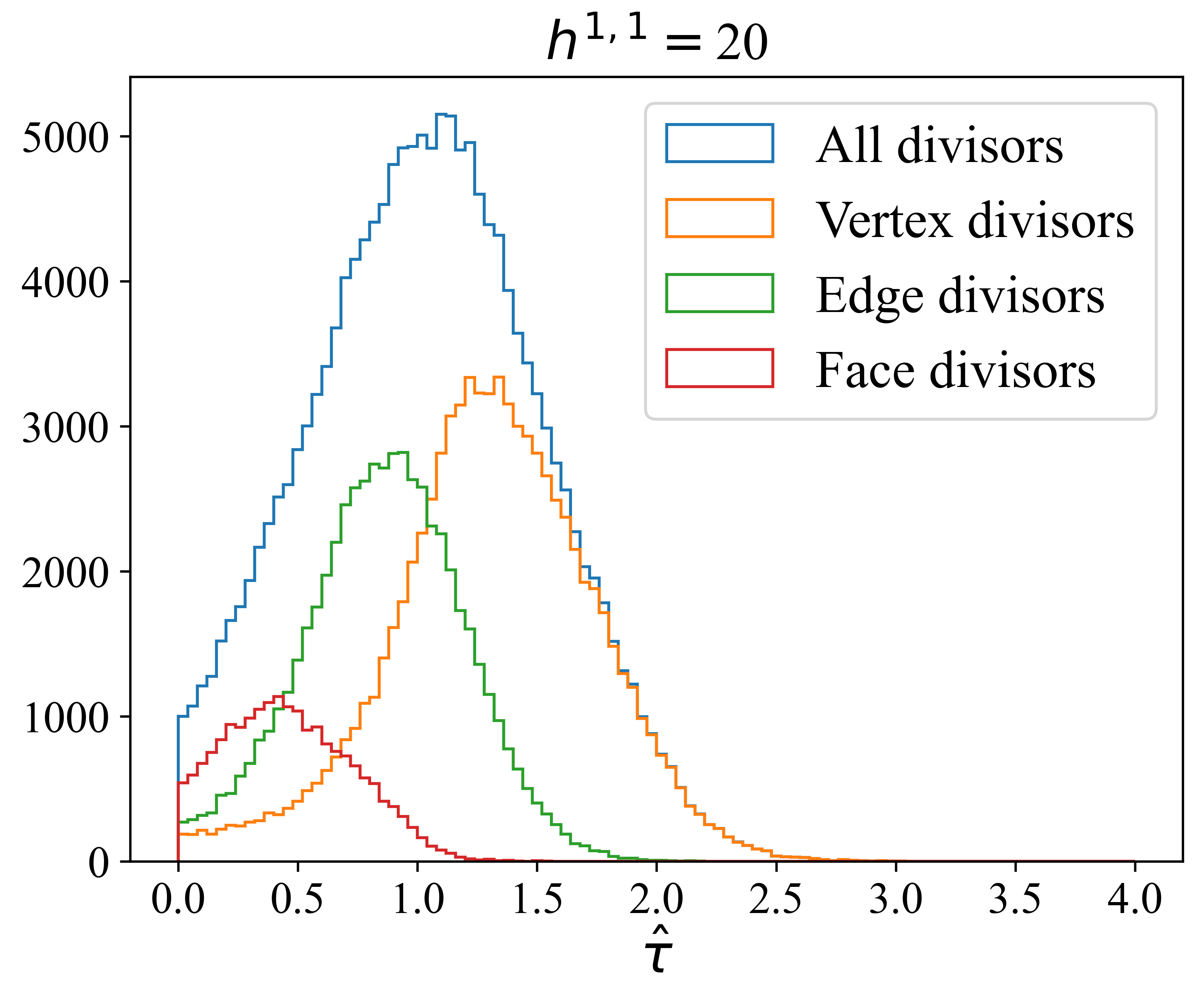}
    \includegraphics[width=0.32\linewidth]{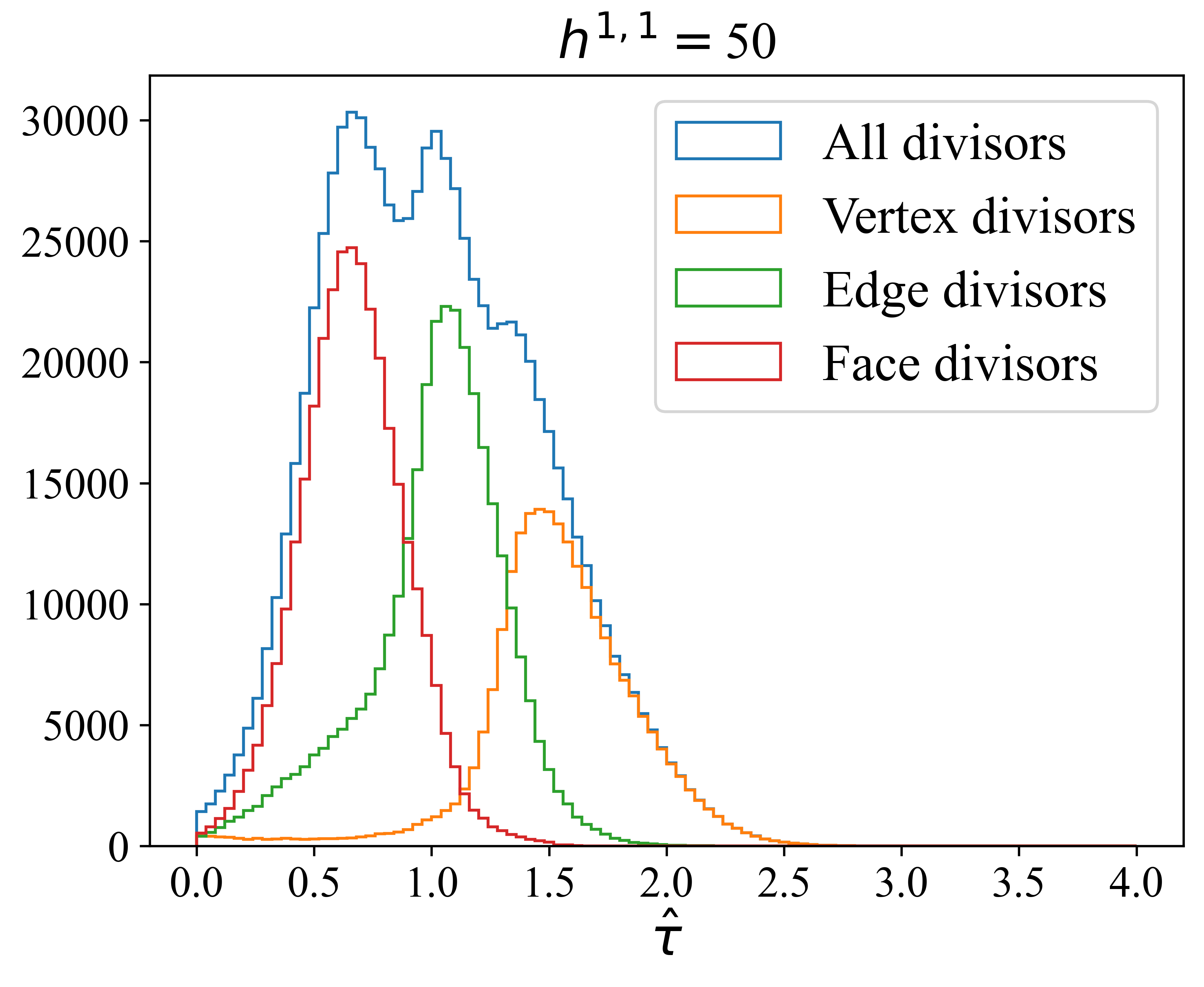}
    \includegraphics[width=0.32\linewidth]{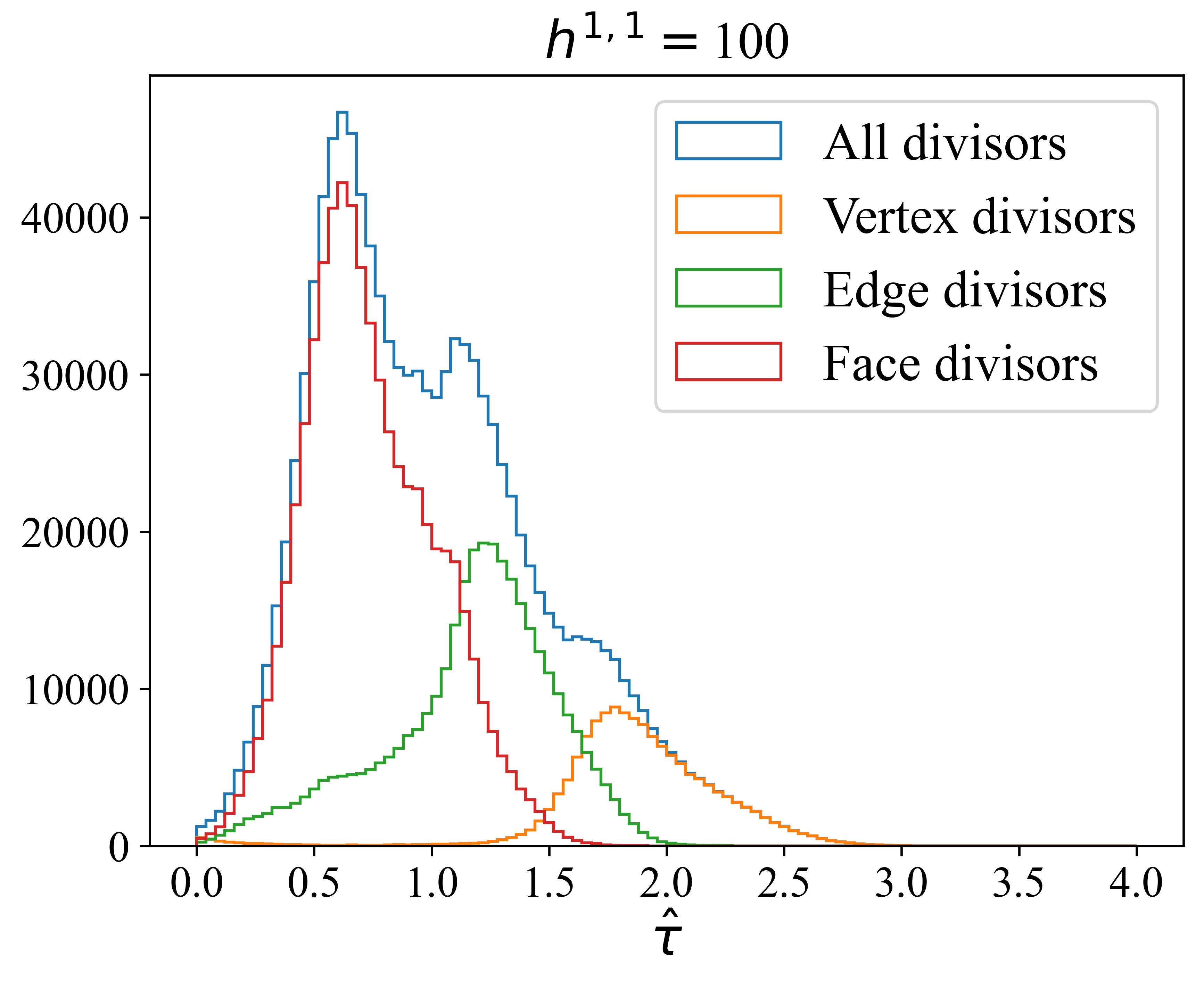}
    \caption{Distributions of different classes of normalized divisor volumes for $h^{1,1}=20,50,100$. We exclude the minimum divisor volume in each model for clarity.}
    \label{fig:tau_vef_normed}
\end{figure}

A prominent feature of Figure~\ref{fig:tau_all} is the appearance of a second peak at around $h^{1,1} \approx 40$. This peak can be easily understood once we analyze the distribution in terms of vertex, edge, and face divisors. Figure~\ref{fig:tau_vef_normed} shows the distributions of $\{\hat{\tau}^I\}$, along with a separation of the divisor volumes into these categories for all Calabi-Yau threefolds in the ensemble $h^{1,1} = 20, \ 50, $ and $100$, respectively. Here we can see that the overall distribution is really composed of three sub-distributions, corresponding to the vertex, edge, and face divisors. At small $h^{1,1}$, there are relatively not many face divisors, and all three distributions sit close together, resulting in a single peak in the overall distribution. As $h^{1,1}$ increases, the number of face divisors increases, and a second peak appears. We expect that for very large $h^{1,1}$ the face divisors will indeed begin to dominate the distribution.

\begin{figure}[h!]
    \centering
    \includegraphics[width=0.49\linewidth]{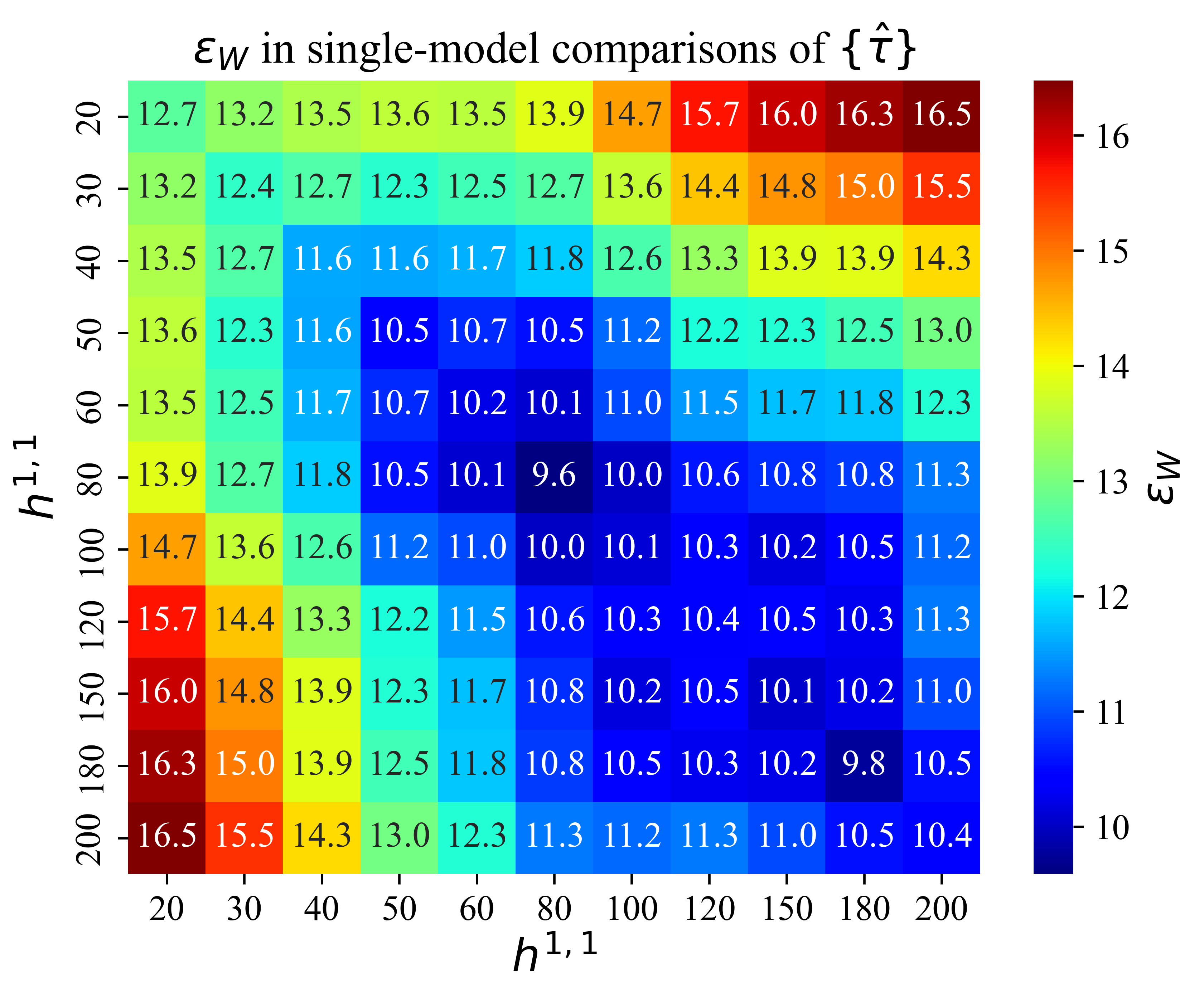}
    \caption{Average $\epsilon_W$ between two sets of $\{\hat{\tau}\}$.}
    \label{fig:KS_percentage_1cy_nontip}
\end{figure}

Our goal is to use the $1$-Wasserstein distance to determine the similarity of distributions of divisor volumes in different Calabi-Yau threefold geometries. We begin by analyzing the overall distributions $\{\hat{\tau}^I\}$ for the Calabi-Yau threefolds in our dataset. We will then analyze the vertex, edge, and face divisor distributions $\{\hat{\tau}^a_{v,e,f}\}$ separately. To this end, we compare Calabi-Yau threefolds across different pairwise combinations of $h^{1,1}$. For each pair of $h^{1,1}$, we randomly select two Calabi-Yau threefolds, $X_1$ and $X_2$, from our ensemble, which is constructed as described in \S\ref{subsec:ensemble}. We then compute $W_1(\{\hat{\tau}^I\}_{X_1}, \{\hat{\tau}^I\}_{X_2})$ for each pair, and subsequently the percentage error in the $1$-Wasserstein distance, $\epsilon_W$. We then repeat this process $1000$  times for each tuple of $h^{1,1}$ values. The results are shown in Figure~\ref{fig:KS_percentage_1cy_nontip}.

We see in Fig.~\ref{fig:KS_percentage_1cy_nontip} that the overall percentage error between individual models is quite low (always below $17\%$), and generally improves as $h^{1,1}$ increases. Geometries with similar $h^{1,1}$ are more likely to have smaller errors: this can also be seen qualitatively in Fig.~\ref{fig:tau_all}, where we can see that the emergence of the second peak in the distribution at moderate $h^{1,1}$ means that models picked from the low-$h^{1,1}$ and high-$h^{1,1}$ range, respectively, will be more dissimilar, resulting in a larger $1$-Wasserstein distance.

\begin{figure}[h!]
    \centering
    \includegraphics[width=0.45\linewidth]{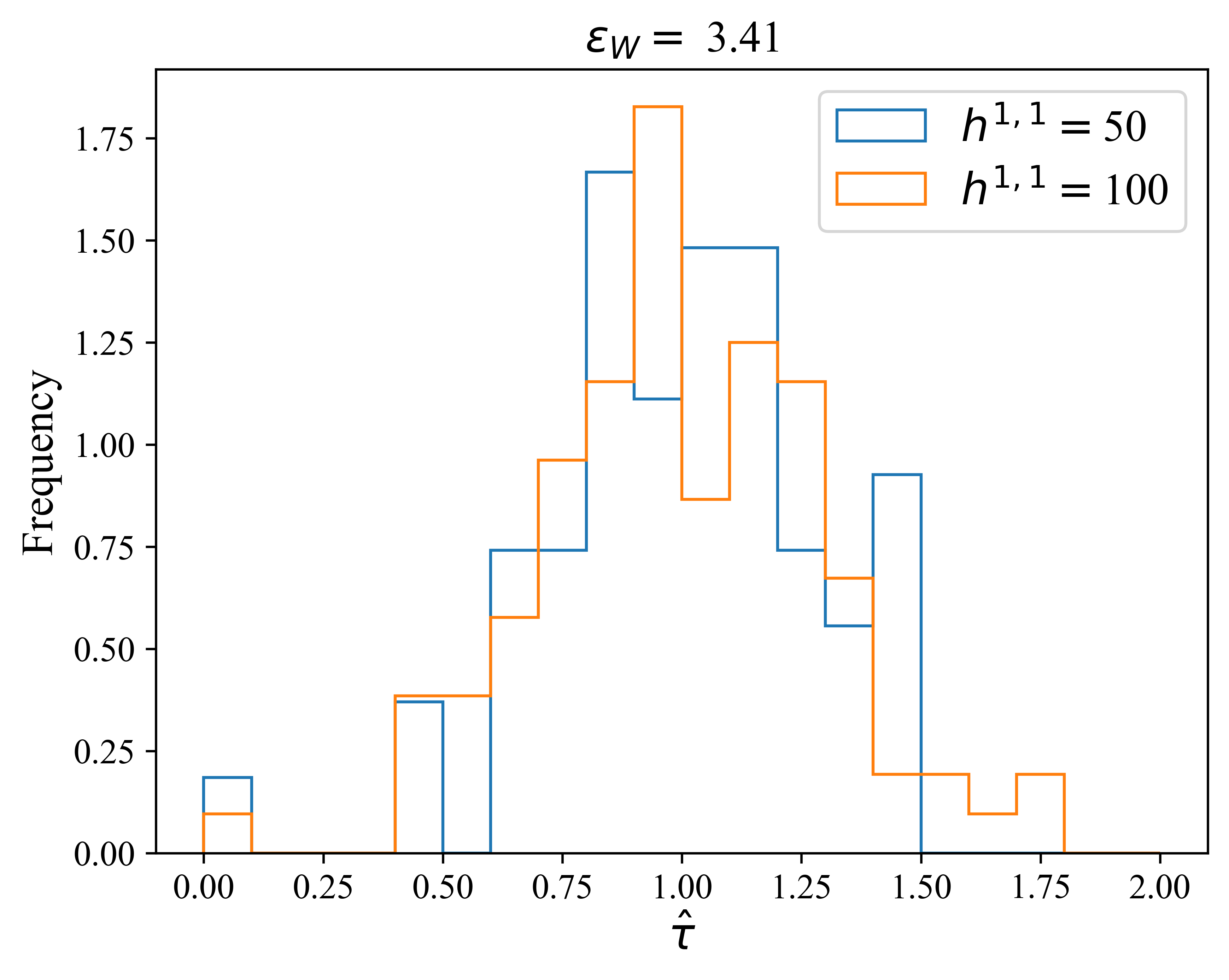}
    \includegraphics[width=0.45\linewidth]{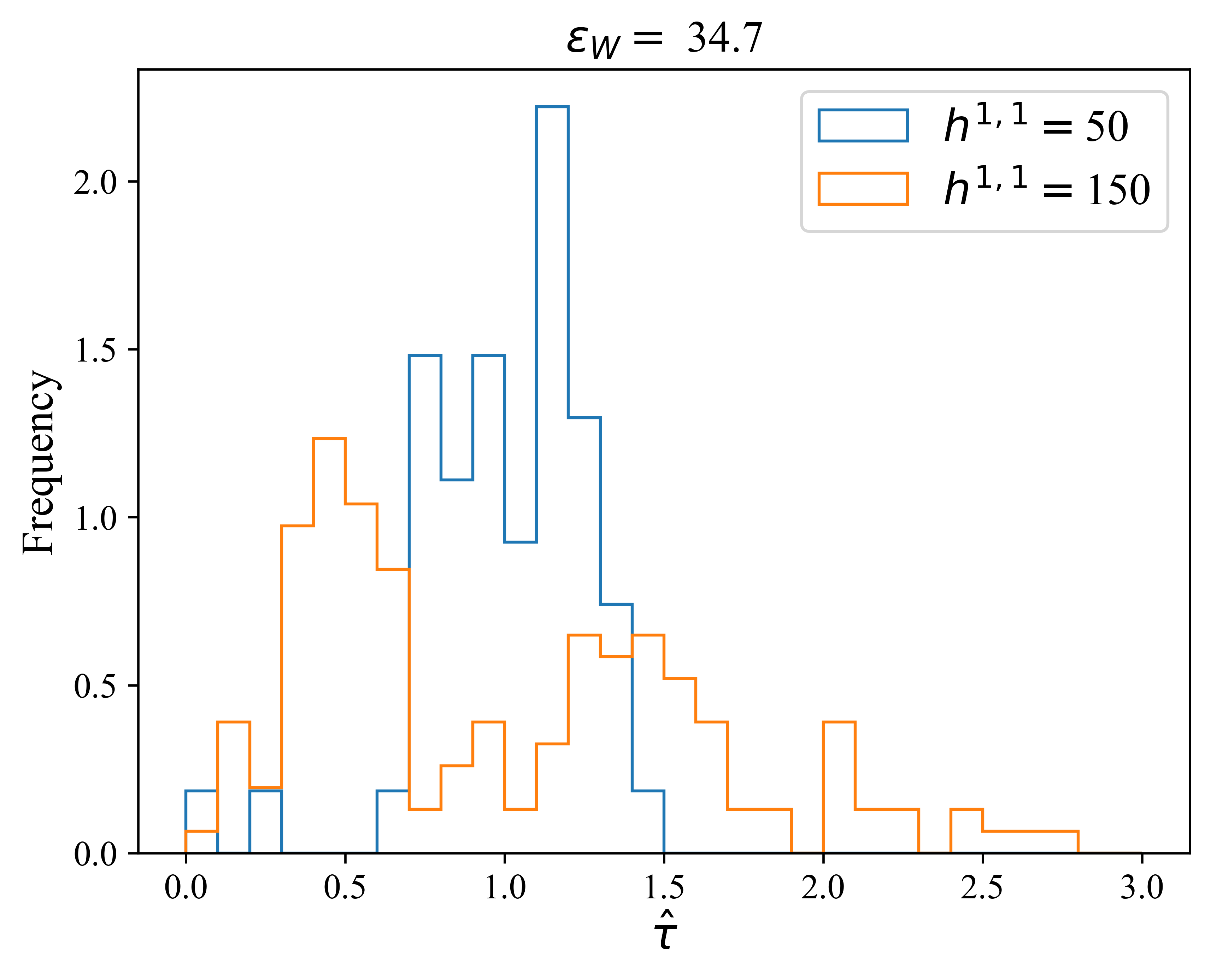}
    \caption{Left: Two sets of $\{\hat{\tau^a}\}$ with similar distributions in distinct Calabi-Yau manifolds. Right: Two sets of $\{\hat{\tau^a}\}$ with different distributions in distinct Calabi-Yau manifolds.
    }
    \label{fig:pairwise_comp}
\end{figure}

To clarify the analysis that the $1$-Wasserstein distance is performing, we pick two representative examples of the pairwise comparisons that it performs and display them in Figure~\ref{fig:pairwise_comp}. The Hodge numbers of these threefolds are different, and so the polytope and triangulation that define the geometries are of course different. The points in the moduli spaces  for each Calabi-Yau threefold are also chosen randomly. On the left, we see two $\{\hat{\tau}^I\}_{X_I}$ (with $h^{1,1}(X_1) =50$ and $h^{1,1}(X_2) = 100$)  distributions that have a percentage error of $\epsilon_W = 3.41$. We can see visually that the distributions are similar. On the right, we show two $\{\hat{\tau}^I\}$ distributions for which $\epsilon_W = 34.7$. In this case, we can see a clear qualitative difference in the distributions: at $h^{1,1} = 150$, there exist two peaks in the distribution, while at $h^{1,1}=50$, there is only a single peak.

\begin{figure}[h!]
    \centering 
    \includegraphics[width=0.32\linewidth]{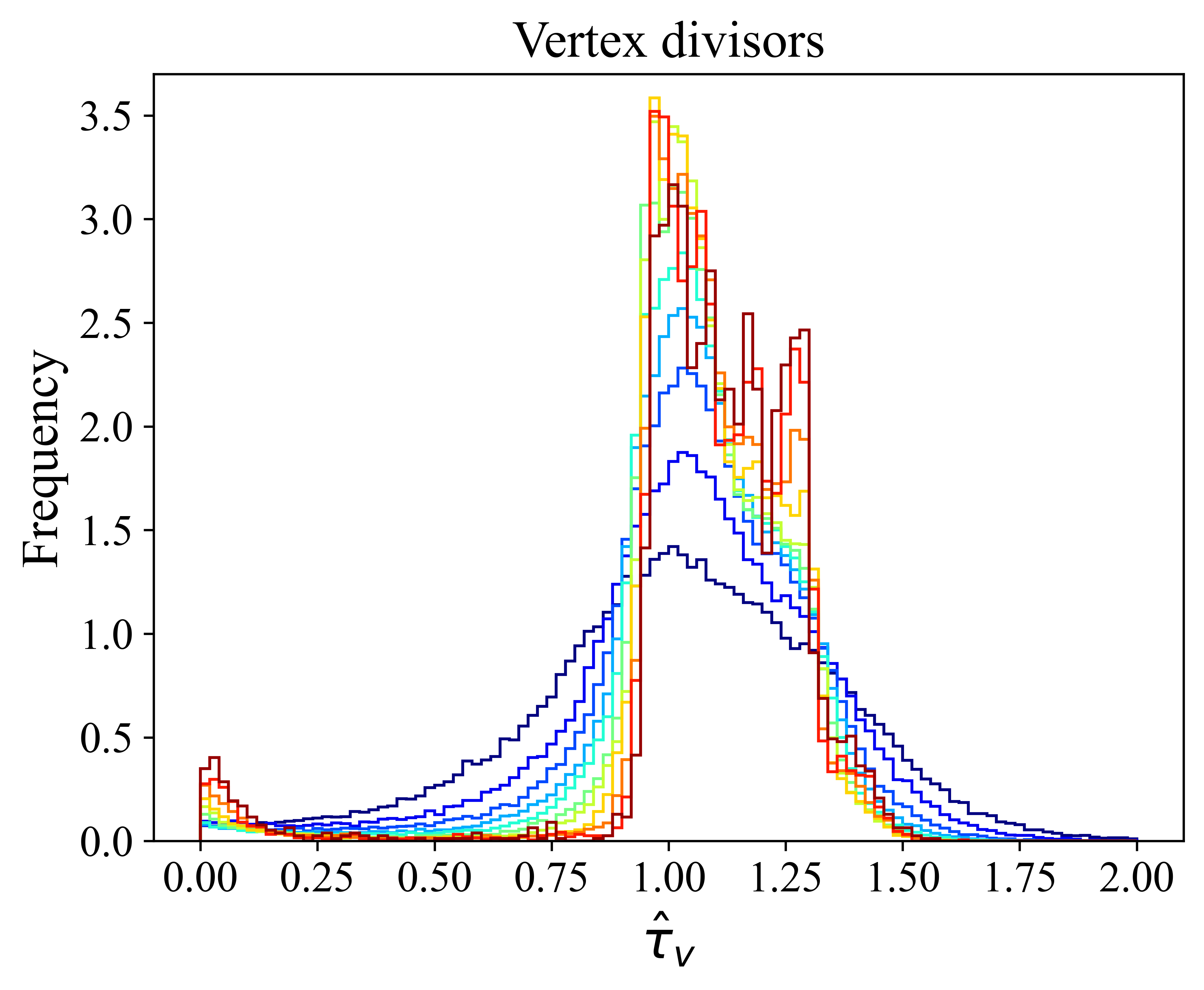}
    \includegraphics[width=0.32\linewidth]{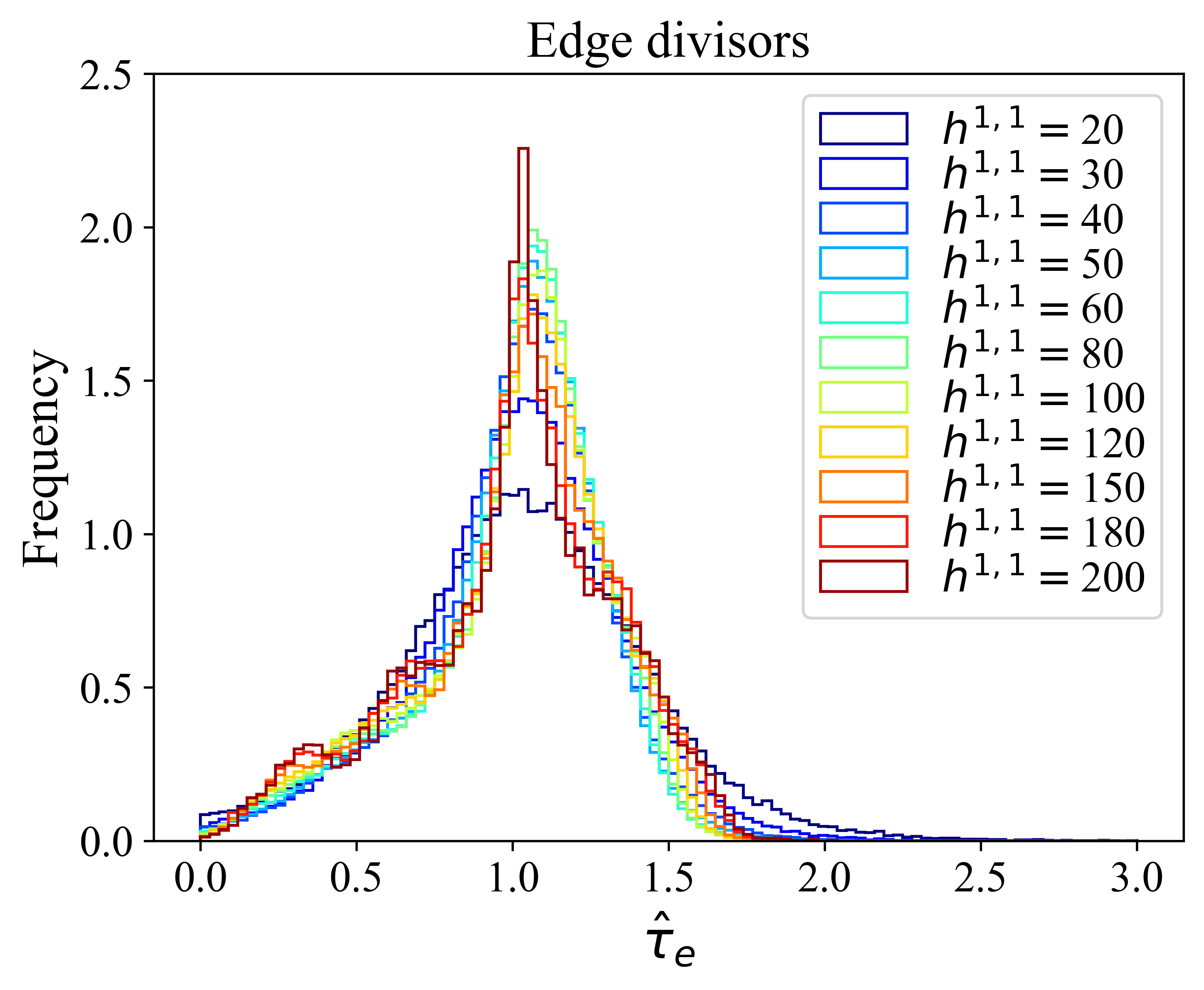}
    \includegraphics[width=0.32\linewidth]{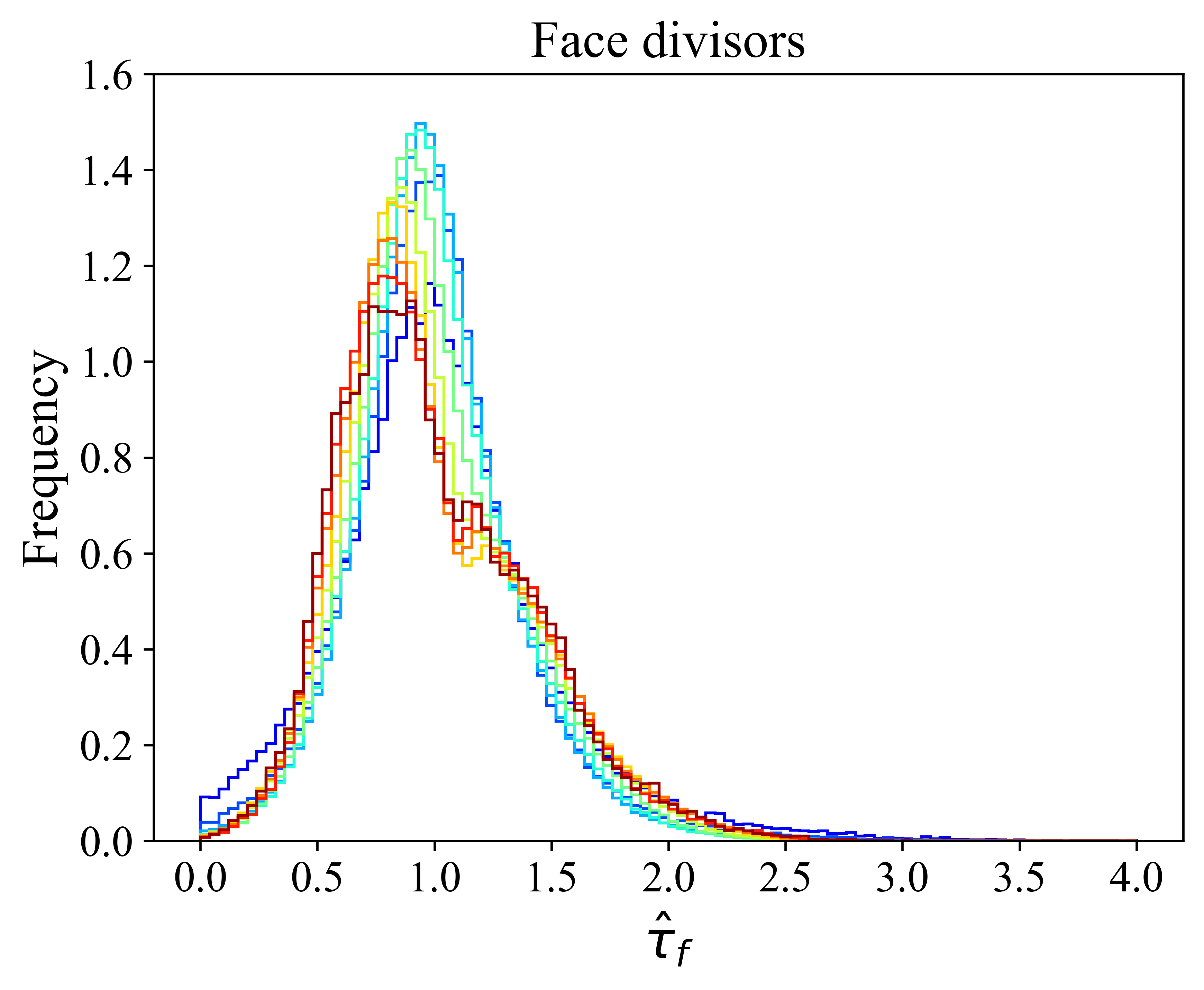}
    \caption{Distributions of the normalized vertex, edge, and face divisor volumes of all the Calabi-Yau threefolds in our dataset, color coded by $h^{1,1}$. The minimum divisor volume of each model has been excluded for clarity.}
    \label{fig:tau_vef_norm}
\end{figure}

\begin{figure}[h!]
    \centering
    \includegraphics[width=0.45\linewidth]{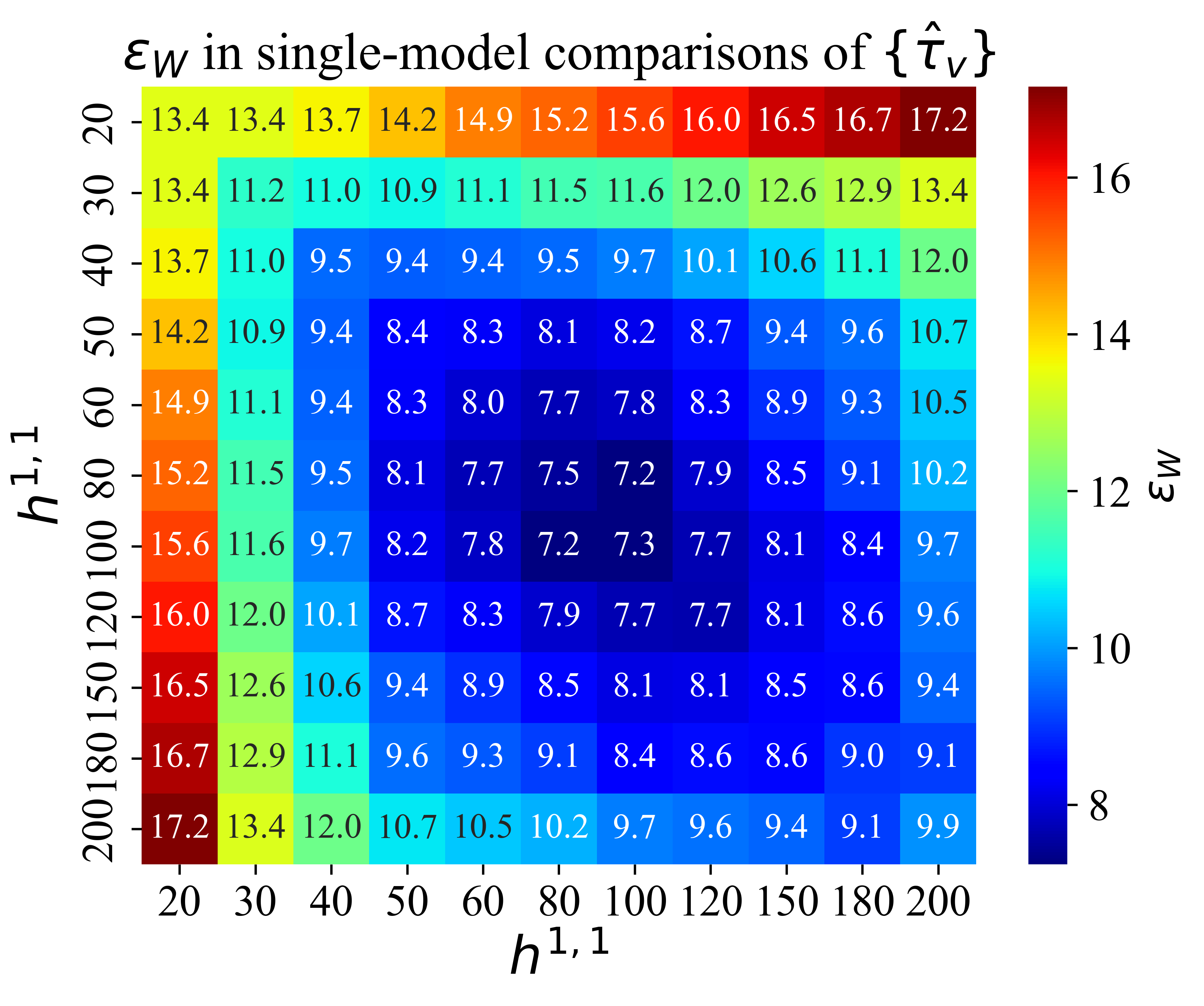}
    \includegraphics[width=0.45\linewidth]{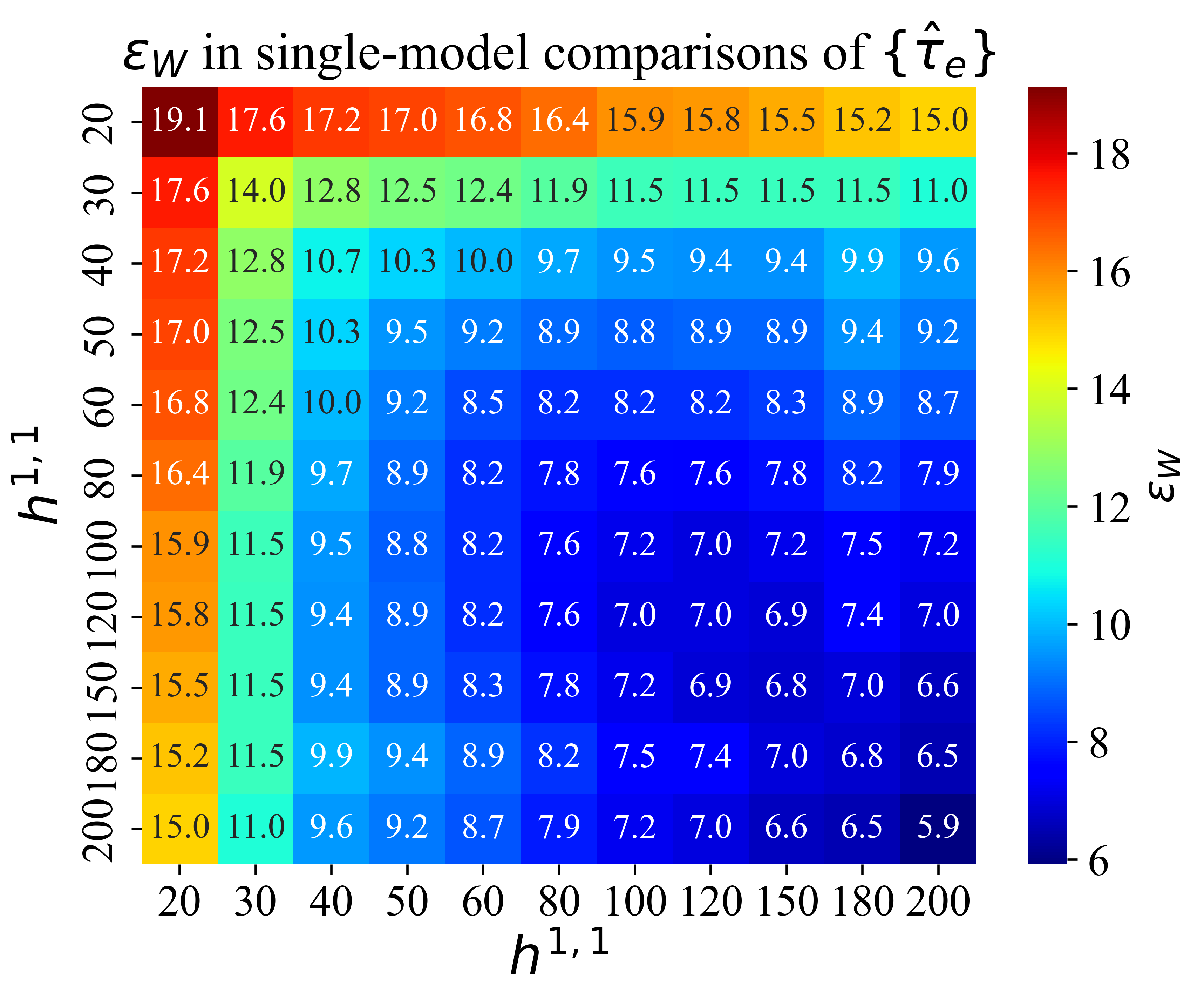}
    \includegraphics[width=0.45\linewidth]{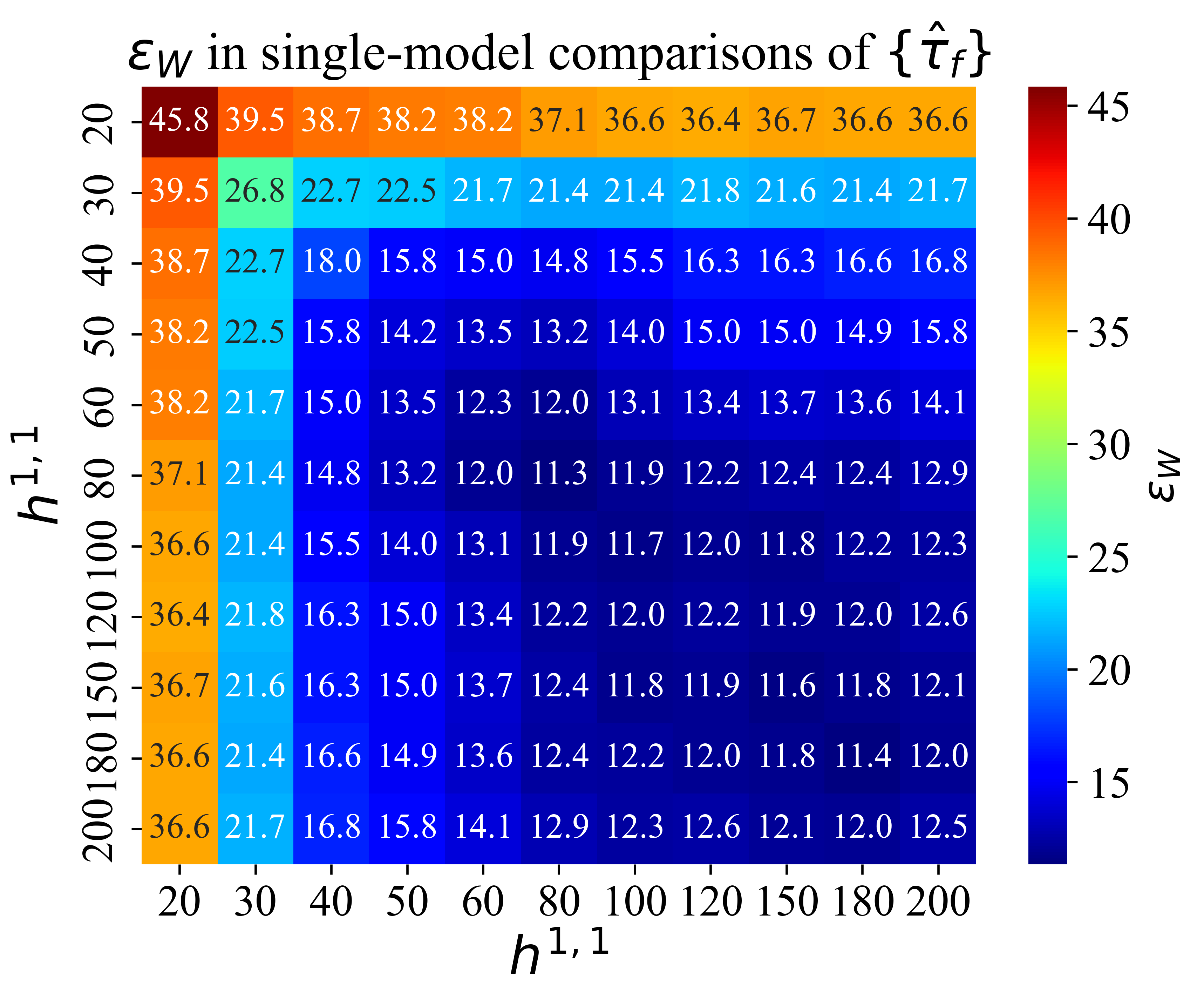}
    \caption{
    Average $\epsilon_W$ between two sets of $\{\hat{\tau}_v\}$, $\{\hat{\tau}_e\}$, and $\{\hat{\tau}_f\}$.}
    \label{fig:KS_percentage_face_1cy_nontip}
\end{figure}

Next, we analyze the vertex, edge, and face divisor distributions separately. Figure~\ref{fig:tau_vef_norm} shows the normalized $\hat{\tau}_v$, $\hat{\tau}_e$, and $\hat{\tau}_f$ data sets from our ensemble. We can see visually that the distributions share similarities both in the bulk and tails. To quantify this, we compute the percentage error in the $1$-Wasserstein distance for pairwise comparisons of geometries among these sub-distributions. Figure~\ref{fig:KS_percentage_face_1cy_nontip} shows the results of the pairwise distance error in our ensemble. Immediately we see similar trends in these distributions to the results for the total divisor distributions in Fig.~\ref{fig:KS_percentage_1cy_nontip}. Again, we see a trend with $h^{1,1}$: the higher overall value $h^{1,1}$, the lower the error $\epsilon_W$. Furthermore, we again see that the larger the gap between the Hodge numbers of the Calabi-Yau threefolds in a given pair, the larger the error is. Note that we observe a significant elevation in the overall scale of the error in the comparison of face divisor distributions with one pair having $h^{1,1}= 20, 30$. This is because at low $h^{1,1}$ there are often \textit{no} face divisors of a given polytope.

In the next section, we will approximate the distribution of divisor volumes with an analytic function, and use this estimation, along with a regression on $\langle \tau^a\rangle$ to model axion observables across the Kreuzer-Skarke landscape.

\section{Modeling the String Axiverse} \label{sec:model}

In this section, we will construct a simplified analytic model for the distribution of divisor volumes in toric hypersurface Calabi-Yau threefolds. We will use the statistical tests described in \S\ref{subsec:stats} to justify this model, and then use it to reconstruct axion observables (such as masses and decay constants) in our ensemble. The goal of this section is to demonstrate that a simple framework for estimating divisor volume distributions is sufficient for approximating axion observables in the Kreuzer-Skarke axiverse, without the need for computationally intensive scans over explicit geometries.

\subsection{Choice of model \label{ssec:which_model}}

Motivated by the conclusion in \S\ref{sec:universal} that normalized divisor volumes $\{\hat{\tau}^I\}$ share common features independent of $h^{1,1}$, we aim to describe the total distribution of divisor volumes as the PDF $p_{\text{fit}}(\hat{\tau})$ of a known analytic distribution. Given such a probability distribution, one can then select $h^{1,1}+4$ independent random variables $\{\hat{\tau}^I\}_{\text{fit}}$ from this distribution to reconstruct a set of normalized divisor volumes. 

We explored a range of candidate analytic probability distributions to model the empirically observed divisor volume distributions across our ensemble of Calabi–Yau threefolds. Among these, we found that the distribution associated with level spacings in the Gaussian Orthogonal Ensemble (GOE) provided the best qualitative and quantitative agreement. Consequently, we adopt the GOE level spacing distribution as our analytic approximation throughout this work. The GOE is a set of real, symmetric matrices
\begin{gather}
    M_n:=\frac{A_n+A_n^T}{\sqrt{2}},
\end{gather}
where all the $n^2$ entries in $A_n$ are independent random numbers following the standard Gaussian distribution \cite{Wigner_1951}. $M_n$ has $n$ real eigenvalues $\lambda_1\leq\lambda_2\leq\cdots\leq\lambda_n$, which famously converge to the Wigner semicircle distribution in the limit of large matrix size $n$,
\begin{align}
    \rho(\lambda/\sqrt{n}) = \frac{1}{2\pi} \sqrt{4-(\lambda/\sqrt{n})^2}.
\end{align}

One can also define the normalized level spacings,
\begin{gather}
    s_i:=\frac{\lambda_{i+1}-\lambda_i}{\frac{1}{n-1}\sum_{i=1}^{n-1}(\lambda_{i+1}-\lambda_i)}.
\end{gather}
These level spacings are known to follow an asymptotic distribution in the large $n$ limit, with a PDF given by
\begin{gather}
    p(s)=\frac{\pi}{2}s \, e^{-\frac{\pi \, s^2}{4}}.
\end{gather}
In this work, we use the GOE level spacing distribution to model normalized divisor volume distributions, i.e we take
\begin{align}
p_{\text{fit}}(\hat{\tau}) = \frac{\pi}{2} \hat{\tau} \, e^{-\frac{\pi \, \hat{\tau}^2}{4}}.
\label{eq:norm_tau_fit}
\end{align}

\begin{figure}
    \centering
    \includegraphics[width=0.6\linewidth]{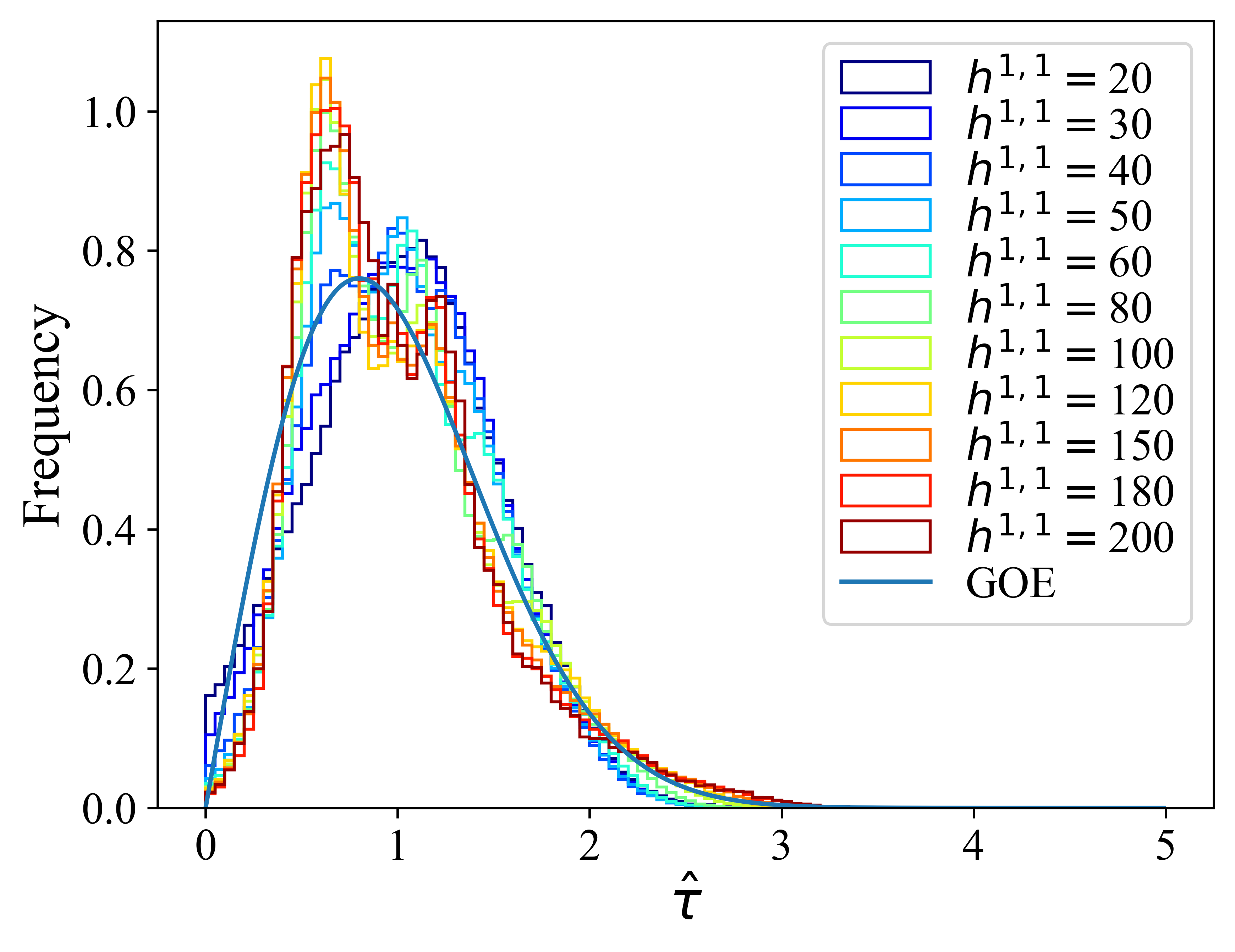}
    \caption{Distributions of normalized divisor volumes color coded by $h^{1,1}$ along with the GOE level spacing PDF.}
    \label{fig:tau_fit}
\end{figure}

\begin{figure}
    \centering
    \includegraphics[width=0.45\linewidth]{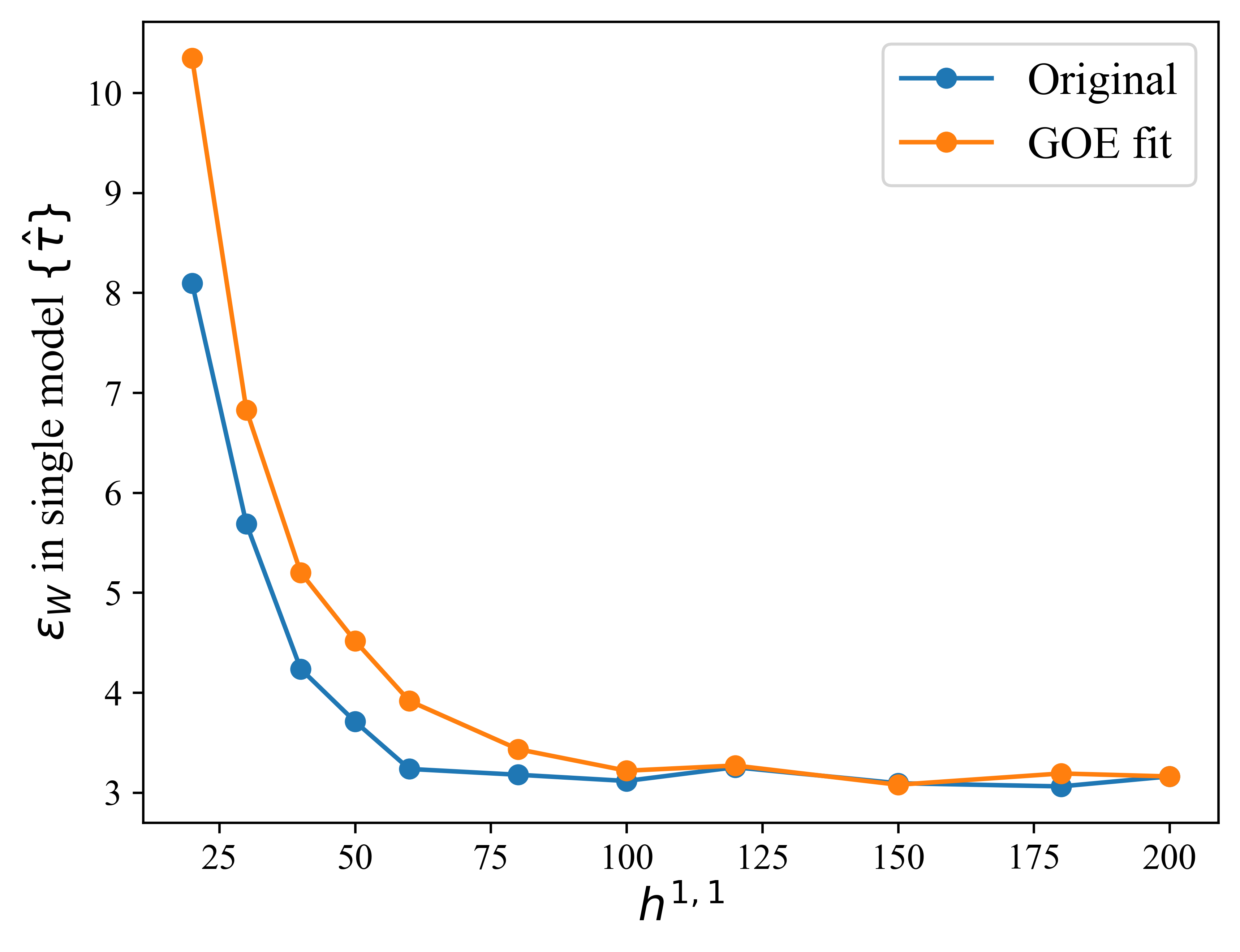}
    \includegraphics[width=0.45\linewidth]{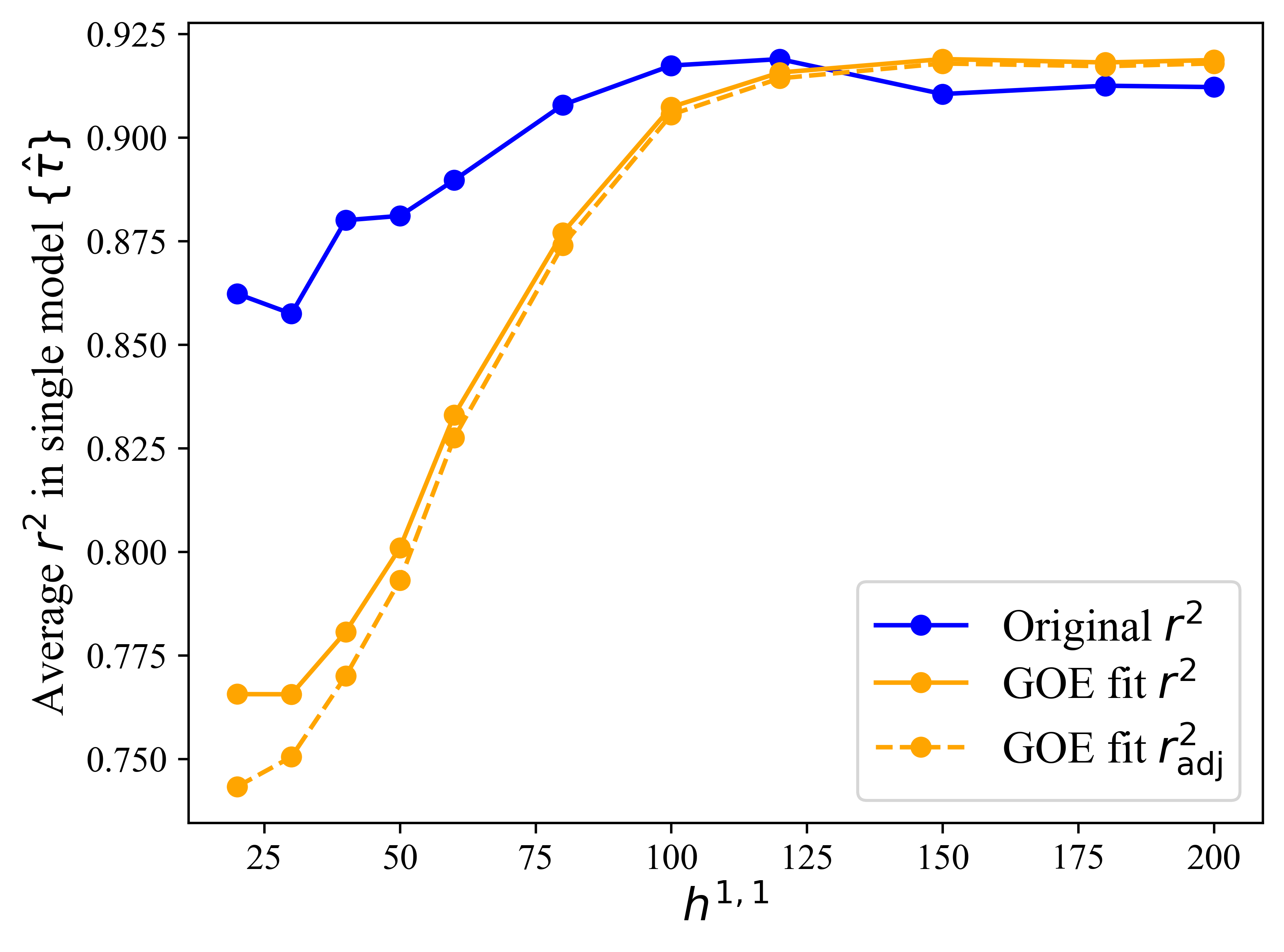}
    \caption{Left: Average percentage error between two models of $\hat{\tau}$. ``Original'' means choosing two random original models and ``GOE fit'' means choosing a random original model and a random model using the GOE fit with the same $h^{1.1}$. Right: Average $r^2$ and adjusted $r^2$ of the same choice of two models. For ``Original'', $r^2$ and $r^2_{\mathrm{adj}}$ are the same by definition.}
    \label{fig:KS_fit_nontip}
\end{figure}

Fig.~\ref{fig:tau_fit} shows $p_{\text{fit}}(\hat{\tau})$ overlaid with the distributions of $\hat{\tau}$ at each value of $h^{1,1}$ in our ensemble. Although $p_{\text{fit}}(\hat{\tau})$ is clearly not a perfect fit for the empirical $\hat{\tau}$ distributions, we find that it constitutes an adequate approximation. To quantify this, we compute the $1$-Wasserstein distances and adjusted $r^2$ values between $p_{\text{fit}}(\hat{\tau})$ and the data. Our comparison method is as follows: for a given trial, we choose one set of $\{\hat{\tau}^I\}_X$ from an explicit Calabi-Yau, $X$, in our ensemble, as well as one set of $\{\hat{\tau}^I\}_{\text{fit}}$ of the same size, drawn randomly from the distribution $p_{\text{fit}}(\hat{\tau})$. We then compute $\epsilon_W$ and $r^2_{\mathrm{adj}}$ between $\{\hat{\tau}^I\}_X$ and $\{\hat{\tau}^I\}_{\text{fit}}$. Fig.~\ref{fig:KS_fit_nontip} shows the average $\epsilon_W$ and $r^2_{\mathrm{adj}}$ values for 1000 trials. As a point of comparison, we also compute the same quantities for random pairs of divisor volume distributions drawn from only the ensemble of explicit geometries, i.e we compare $\{\hat{\tau}^I\}_{X_1}$ and $\{\hat{\tau}^I\}_{X_2}$ for random pairs of geometries, $X_1$ and $X_2$. This comparison conveys the size of fluctuations within the empirical ensemble. We can see that as $h^{1,1}$ increases, the percentage error in the $1$-Wasserstein distance decreases, while the adjusted $r^2$ value increases. This signals that our model $p_{\text{fit}}(\hat{\tau})$ performs better as $h^{1,1}$ increases, which is what one would expect.

\begin{figure}[h!]
    \centering
    \includegraphics[width=0.45\linewidth]{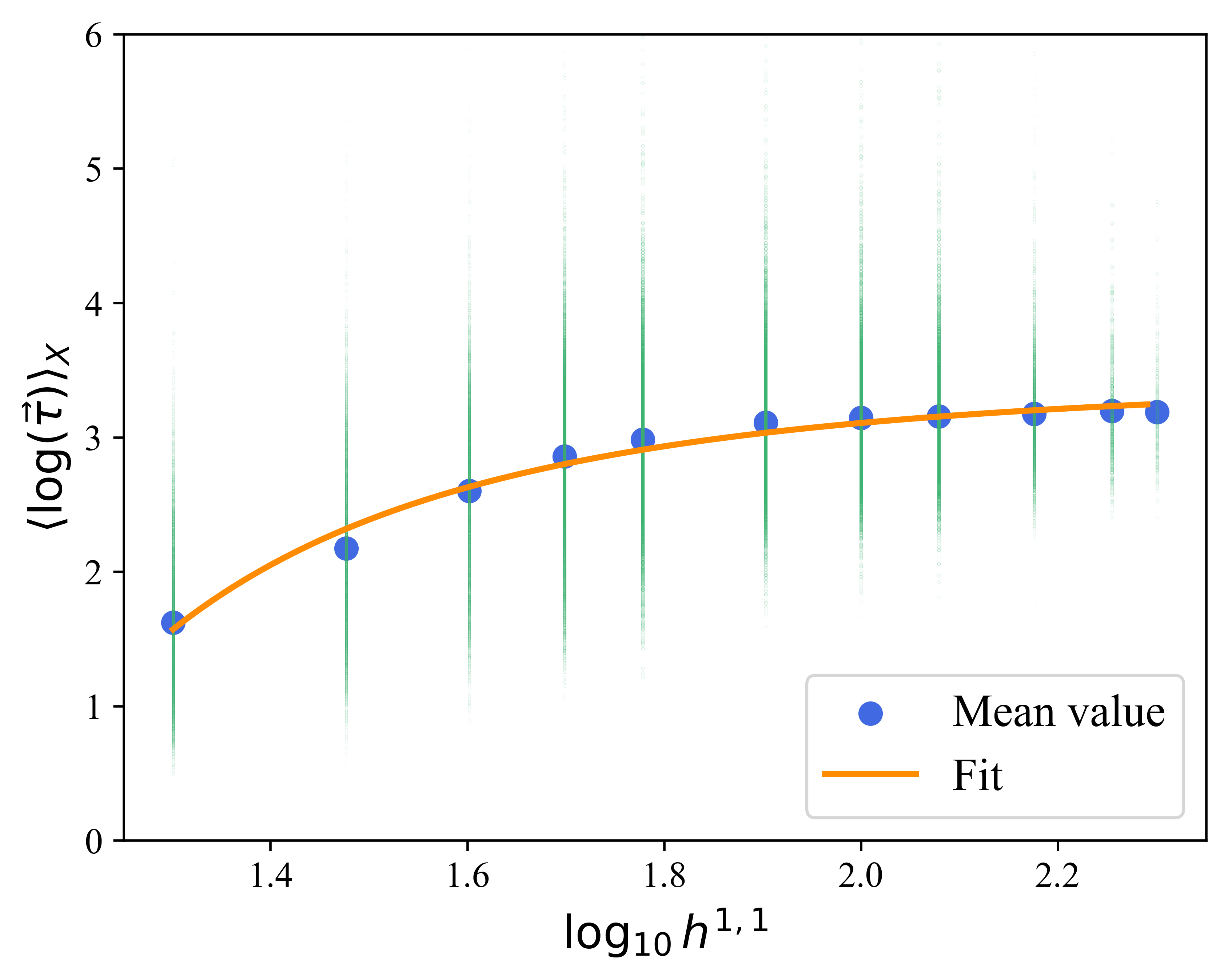}
    \caption{Scatter plot of $\langle\log(\vec{\tau})\rangle_X$ as a function of $\log(h^{1,1})$. The mean values of the scatter for each value of $h^{1,1}$ are shown in blue, while our best fit is shown in orange.
    }
    \label{fig:tau_fit_mean}
\end{figure}

The analytic distribution $p_{\text{fit}}(\hat{\tau})$ produces a good approximation of the normalized divisor volume distributions, but to model the Kreuzer-Skarke axiverse as a function of $h^{1,1}$, we have to introduce information about the \textit{mean} of the divisor volume distributions, which increases as a function of $h^{1,1}$. Fig.~\ref{fig:tau_fit_mean} shows the scatter and average value of $\langle \text{log}(\vec{\tau})\rangle_X$ as a function of $h^{1,1}$ in our ensemble. We fit this trend as
\begin{gather}
    \label{eq:mean_fit}
    \langle \log(\vec{\tau})\rangle_{\text{fit}}=a-b\times(\log_{10}h^{1,1})^{-c},
\end{gather}
where $a\approx 3.433$, $b \approx 5.404$, and $c \approx 4.050$ are fit parameters. Our procedure to reconstruct a distribution of divisor volumes at a particular value of $h^{1,1}$ is to first construct the normalized divisor volumes by obtaining $h^{1,1}+4$ random draws from the probability distribution \eqref{eq:norm_tau_fit}. Then, using the fit \eqref{eq:mean_fit}, we construct a set of fit divisor volumes as
\begin{align}
    \log(\tau_{\text{fit}}^I) = \hat{\tau}_{\text{fit}}^I \times \langle \log(\vec{\tau})\rangle_{\text{fit}}.
    \label{eq:tau_fit}
\end{align}

\begin{figure}
    \centering
    \includegraphics[width=0.45\linewidth]{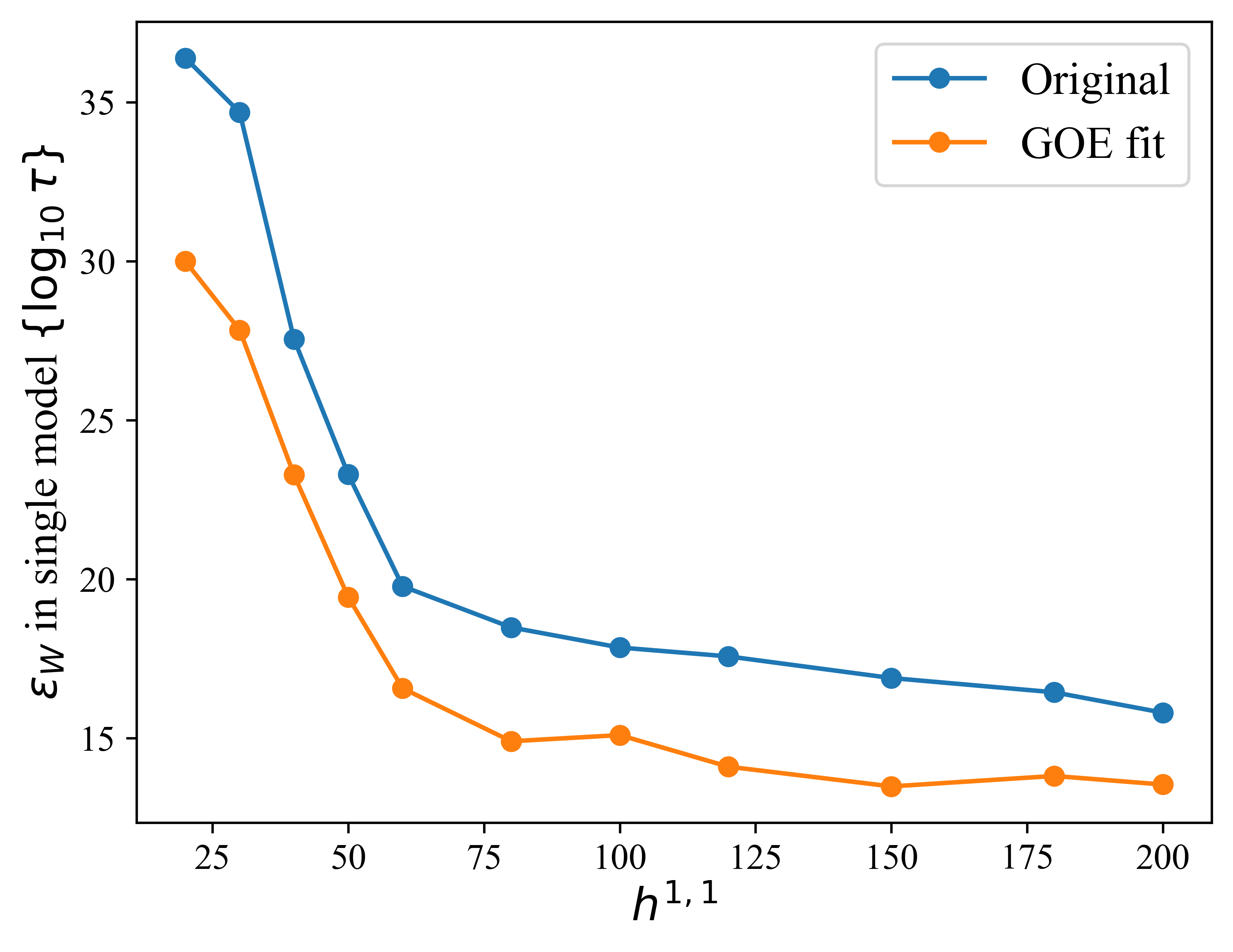}
    \includegraphics[width=0.45\linewidth]{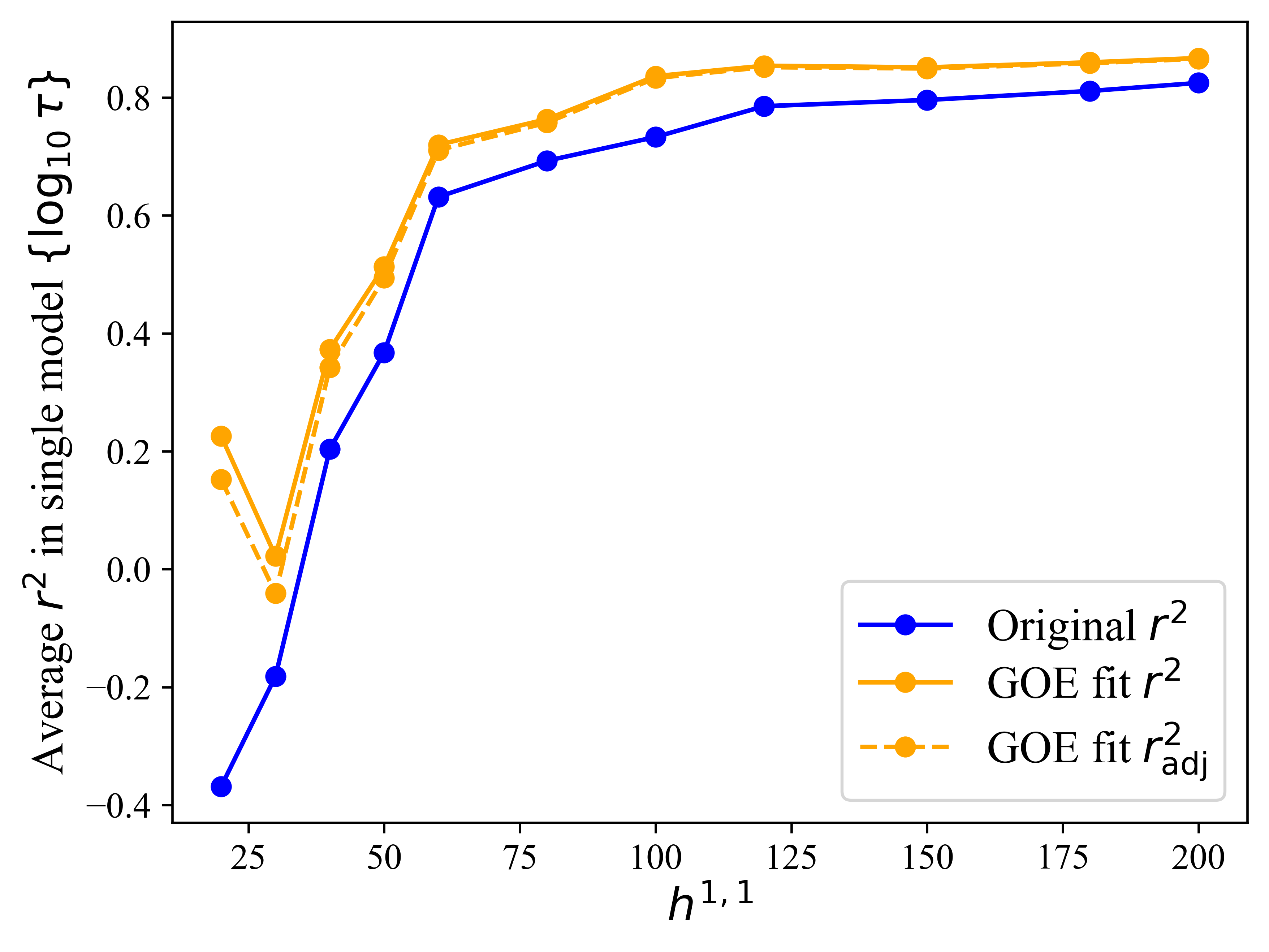}
    \caption{Left: Average percentage error between two models of $\log(\tau)$. Right: Average $r^2$ and adjusted $r^2$ of the same choice of two models.}
    \label{fig:KS_fit_tau}
\end{figure}

Again, we assess the quality of the model by computing $\epsilon_W$ for individual models in our ensemble compared to divisor volume distributions reconstructed using \eqref{eq:tau_fit}, as well as by computing the $r^2_{\mathrm{adj}}$ value in these same pairwise comparisons. The results are shown in Fig.~\ref{fig:KS_fit_tau}. Here we can see again that the model \eqref{eq:tau_fit} performs better as $h^{1,1}$ increases, with the percentage error in the $1$-Wasserstein distance plateauing under $15\%$.

We also assessed a version of the model where instead of fitting the overall divisor volume distribution to an analytic PDF, we modeled the vertex, edge, and face divisors separately. The motivation for this model is that the distributions $\{\hat{\tau}_{v,e,f}\}$ (as defined in \eqref{eq:normed_tau_v}) exhibit somewhat more universal properties, as shown in \S\ref{sec:universal}. The downside of a model like this is that because it involves modeling three separate distributions as a function of $h^{1,1}$, it involves multiplicatively more fit parameters. Indeed, using this model we find typically slightly lower $\epsilon_W$, but worse $r^2_{\mathrm{adj}}$ values, indicating that the model is penalized for having more parameters. We therefore use the fit \eqref{eq:tau_fit} to model overall divisor volume distributions. In the next subsection, we will use this model to reproduce axion observables in the Kreuzer-Skarke landscape.

\subsection{Reproducing axion observables \label{ssec:mf}}

Finally armed with a model of divisor volumes \eqref{eq:tau_fit}, we can compute axion related physical observables as described in \S\ref{subsec:axEFTs}. This section focuses on the axion masses $m_a$ and decay constants $f_a$ using the approximations \eqref{eq:m_approx} and \eqref{eq:f_approx}. 

\begin{figure}[h!]
    \centering
    \includegraphics[width=\linewidth]{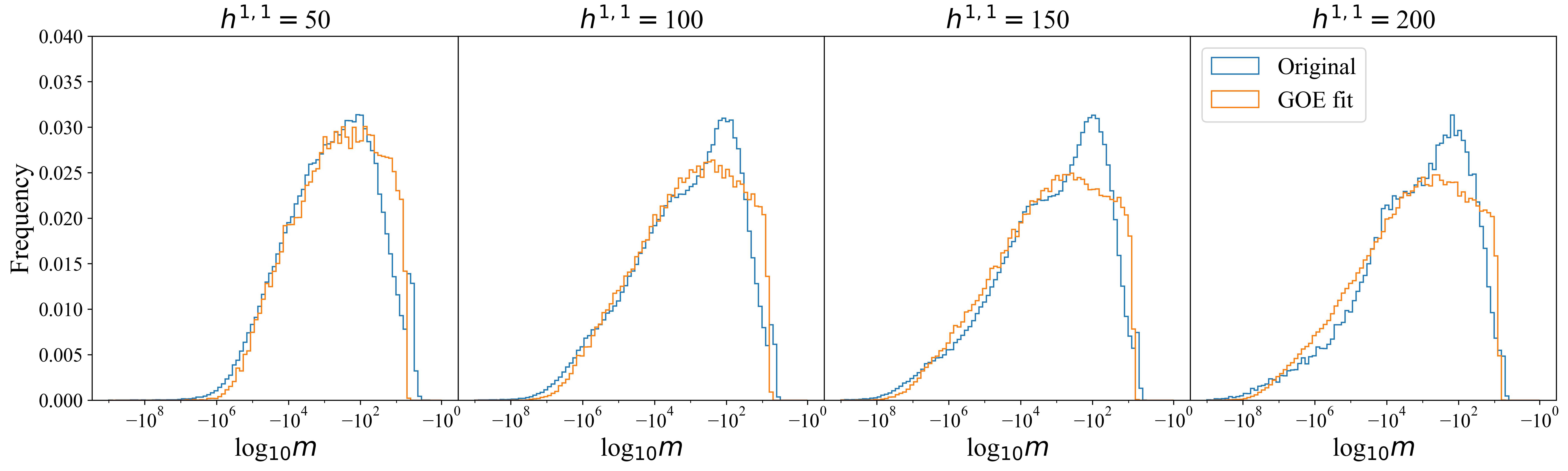}
    \caption{Distributions of axion masses computed using $\{\tau^a\}$ and $\{\tau^a_{\text{fit}}\}$.}
    \label{fig:m_distribution}
\end{figure}

\begin{figure}
    \centering
    \includegraphics[width=\linewidth]{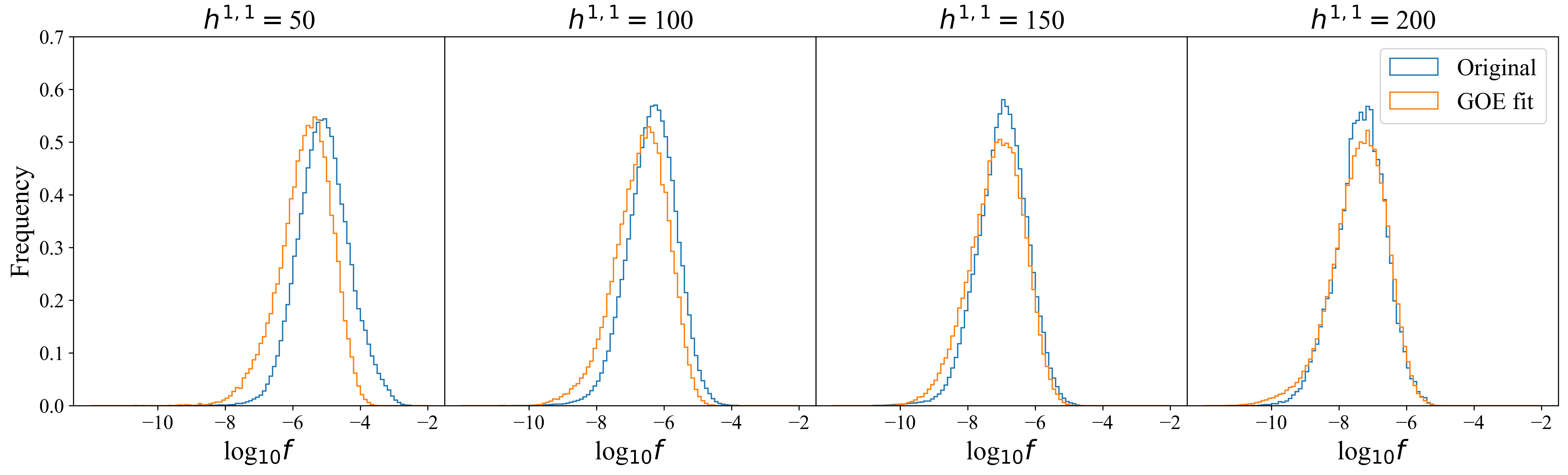}
    \caption{Distributions of axion decay constants computed using $\{\tau^a\}$ and $\{\tau^a_{\text{fit}}\}$.}
    \label{fig:f_distribution}
\end{figure}

To model axion observables at a given $h^{1,1}$, we begin by constructing a set of normalized divisor volumes using our GOE model \eqref{eq:norm_tau_fit}. Then, we multiply the normalized divisor volumes by the mean obtained from the fit \eqref{eq:mean_fit}. Finally, we use the resulting divisor volumes as input in the approximations \eqref{eq:m_approx} and \eqref{eq:f_approx} to obtain a fitted set of axion masses and decay constants. Note that once we have specified our model, there is no further input needed from the explicit ensemble to obtain approximations for the axion observables. Fig.~\ref{fig:m_distribution} shows the true versus approximated mass distributions for several representative values of $h^{1,1}$, while Fig.~\ref{fig:f_distribution} shows the comparison of decay constant distributions. Our model clearly reproduces these axion observables quite well.


\begin{figure}[h!]
    \centering
    \includegraphics[width=0.45\linewidth]{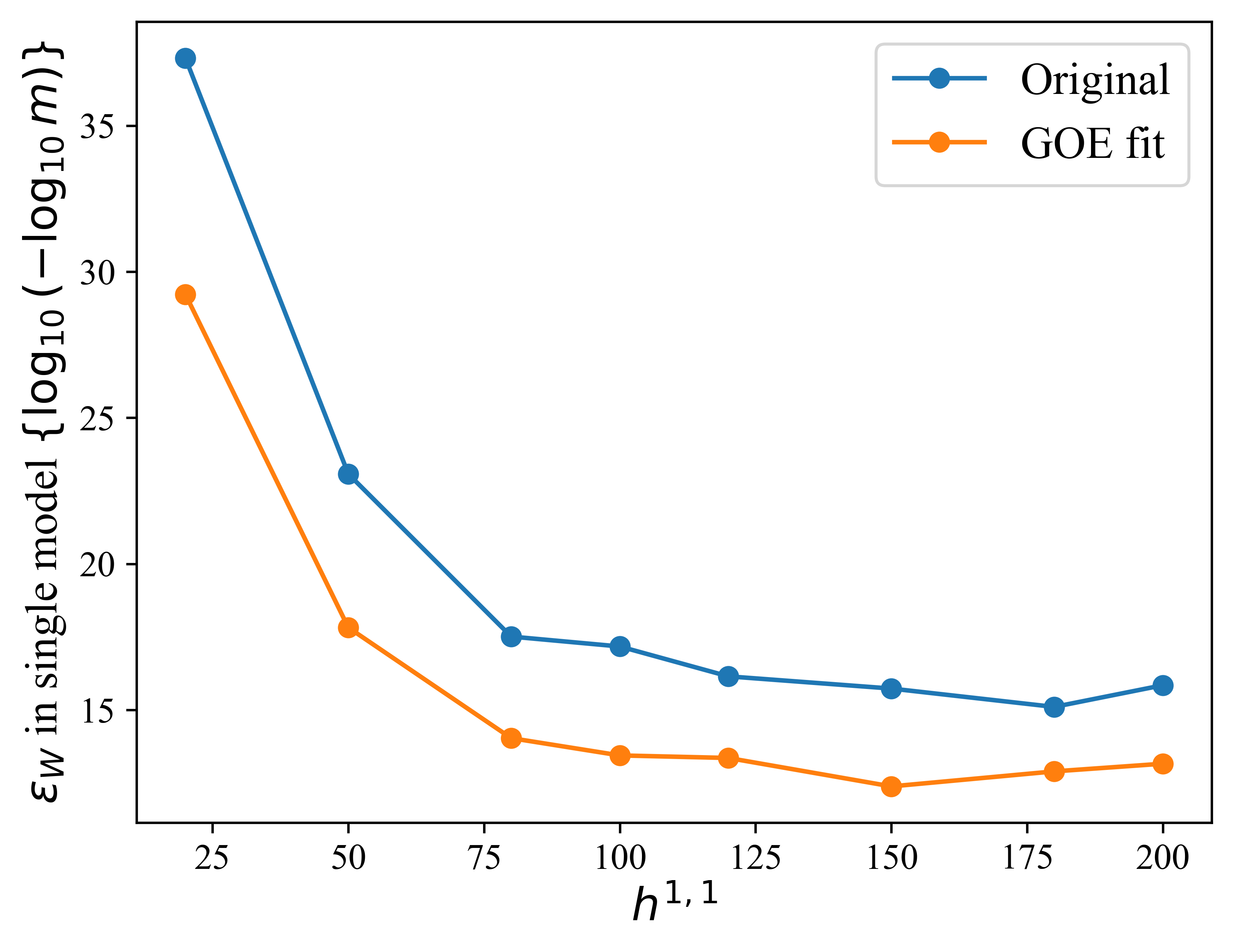}
    \includegraphics[width=0.45\linewidth]{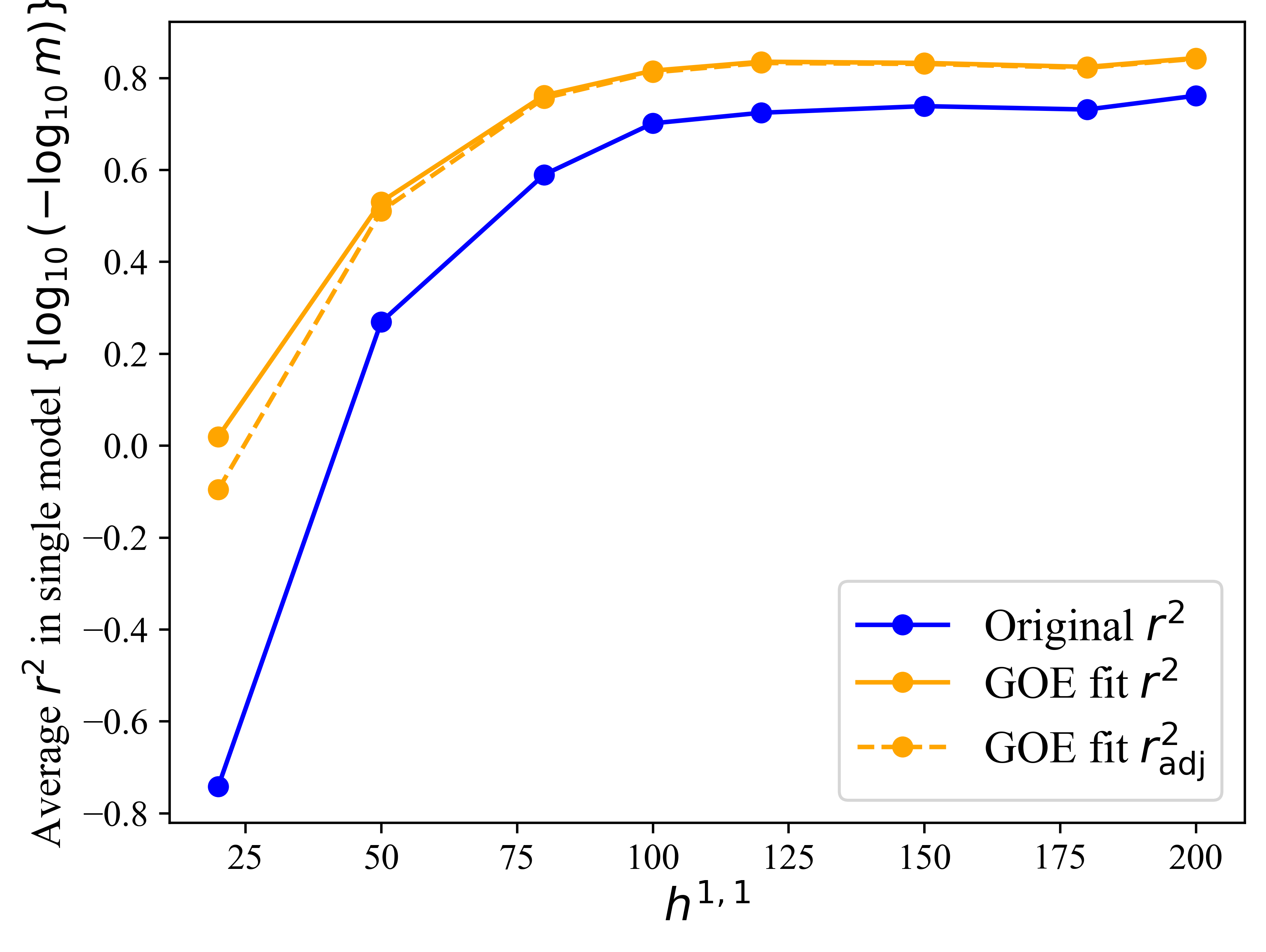}
    \caption{Left: Average percentage error between two models of $\log(-\log(m))$. Right: Average $r^2$ and adjusted $r^2$ of the same choice of two models.}
    \label{fig:KS_fit_m}
\end{figure}

\begin{figure}[h!]
    \centering
    \includegraphics[width=0.45\linewidth]{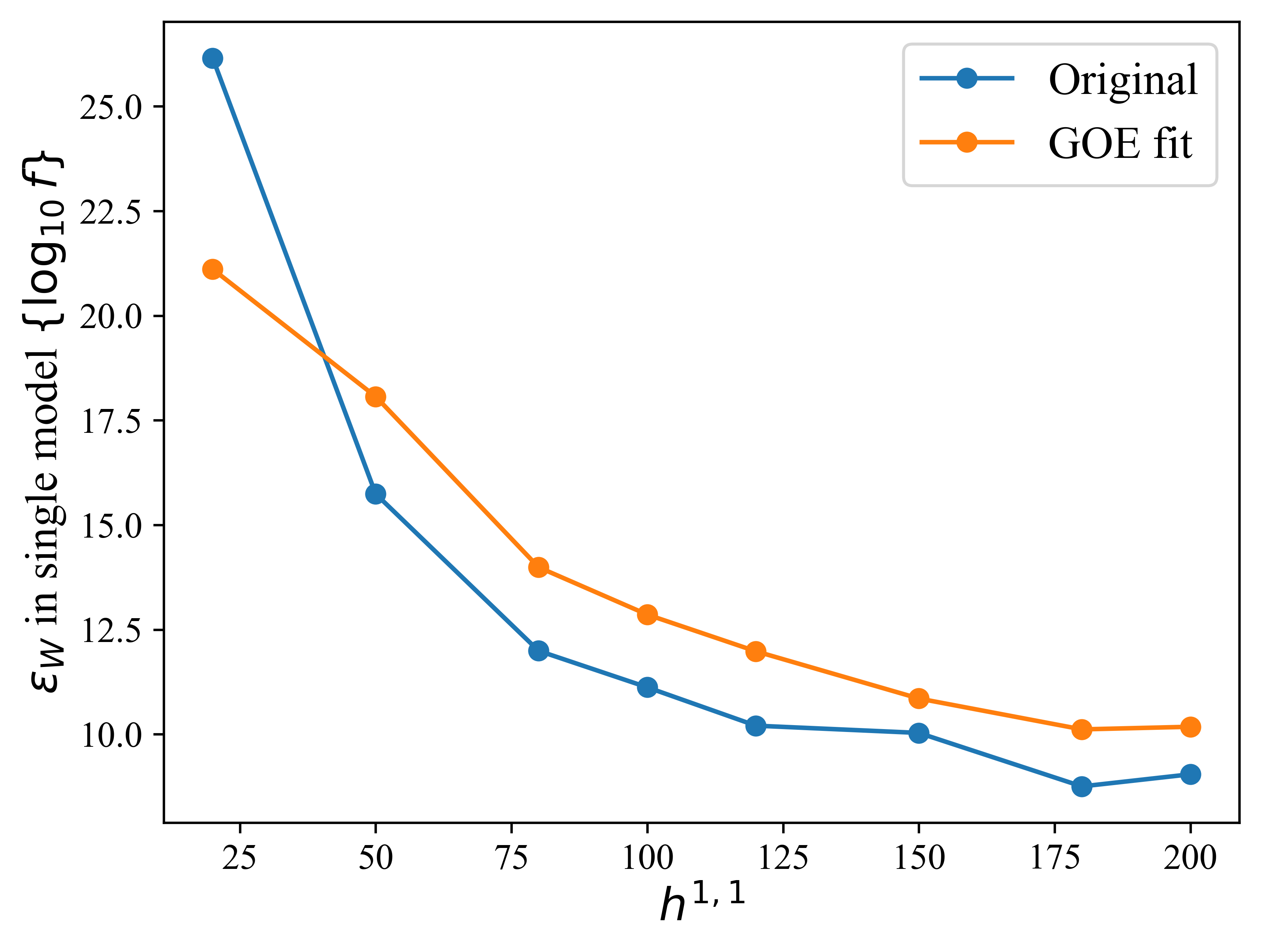}
    \includegraphics[width=0.45\linewidth]{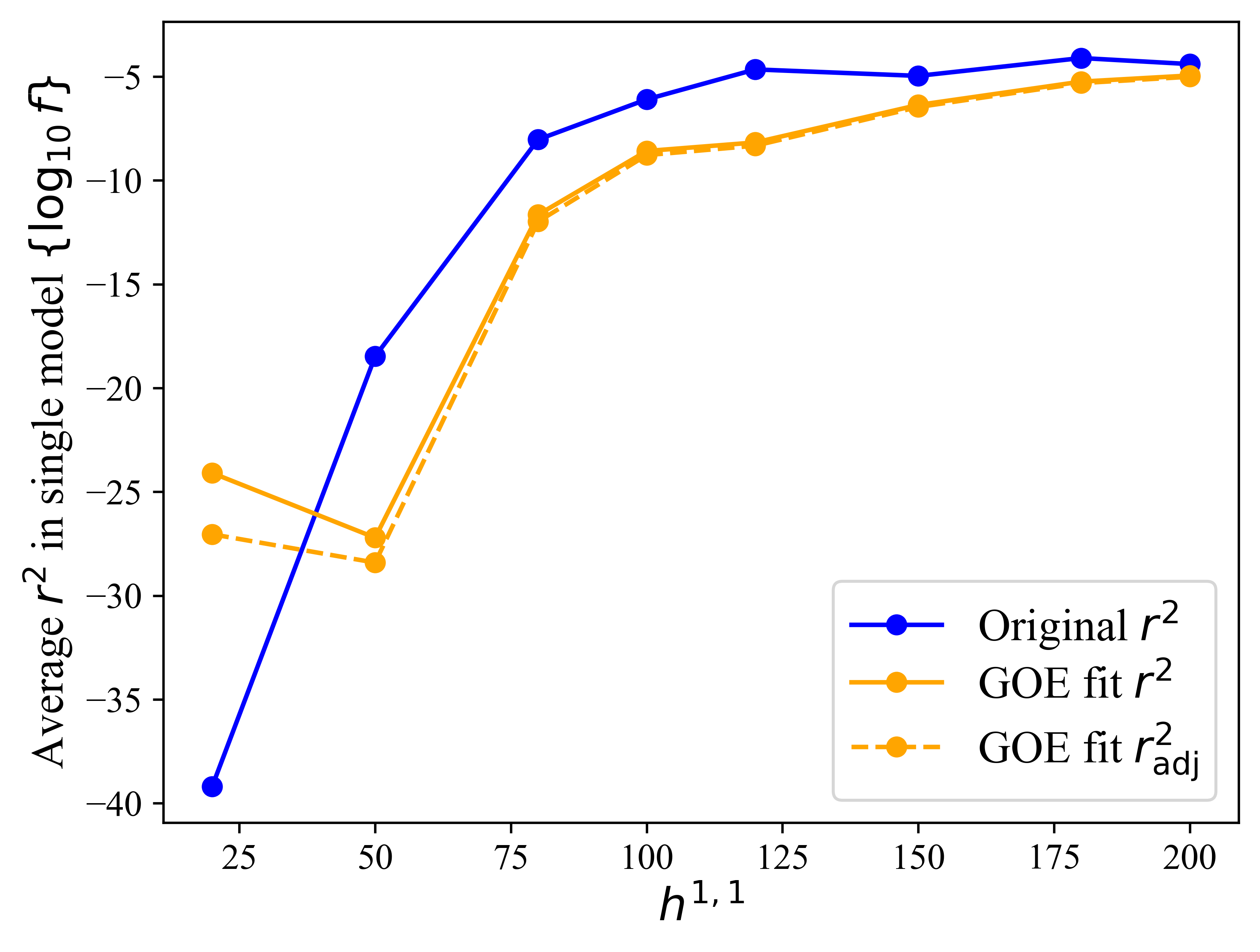}
    \caption{Left: Average percentage error between two models of $\log(f)$. Right: Average $r^2$ and adjusted $r^2$ of the same choice of two models. 
    }
    \label{fig:KS_fit_f}
\end{figure}

To quantify this statement, we compare the predicted observables using our model with the exact quantities computed as in \S\ref{subsec:axEFTs}. As in previous sections, to assess the efficacy of this model, we compare true masses and decay constants with the reconstructed ones using the $1$-Wasserstein distance and the adjusted $r^2$ value. In particular, we choose a random geometry $X$ from our explicit ensemble, compute the true axion decay constants $\{f_i\}_X$ and masses $\{m_i\}_X$ using \eqref{eq:decayconst} and \eqref{eq:mass}. Then, we perform a random draw from our model \eqref{eq:tau_fit} of size $h^{1,1}+4$. We then compute the approximate decay constants $\{f_i\}_{\text{fit}}$ and masses $\{m_i\}_{\text{fit}}$ using \eqref{eq:f_approx} and \eqref{eq:m_approx}. The distributions $\{f_i\}_X$ and $\{f_i\}_{\text{fit}}$ (respectively, $\{m_i\}_X$ and $\{m_i\}_{\text{fit}}$) are compared by computing $\epsilon_W$ and $r^2_{\mathrm{adj}}$.  Fig.~\ref{fig:KS_fit_m} shows these values as a function of $h^{1,1}$. The trend indicates that the model performs better at reproducing masses as a function of $h^{1,1}$. Similarly, Fig.~\ref{fig:KS_fit_f} shows the percentage error in the $1$-Wasserstein distance and adjusted $r^2$ values for decay constant distributions. As explained in \S\ref{subsec:divapprox}, the percentage errors for the decay constant distributions are lower than those for the masses because we are comparing masses on a doubly log scale. Additionally, we note that the $r^{2}$ values for single-model comparisons within both the original ensemble and approximated distributions are negative. This is due to the fact that fluctuations in decay constants from model to model are large. We therefore focus only on the relative values between the original dataset and the GOE fit.

\begin{figure}
    \centering
    \includegraphics[width=0.45\linewidth]{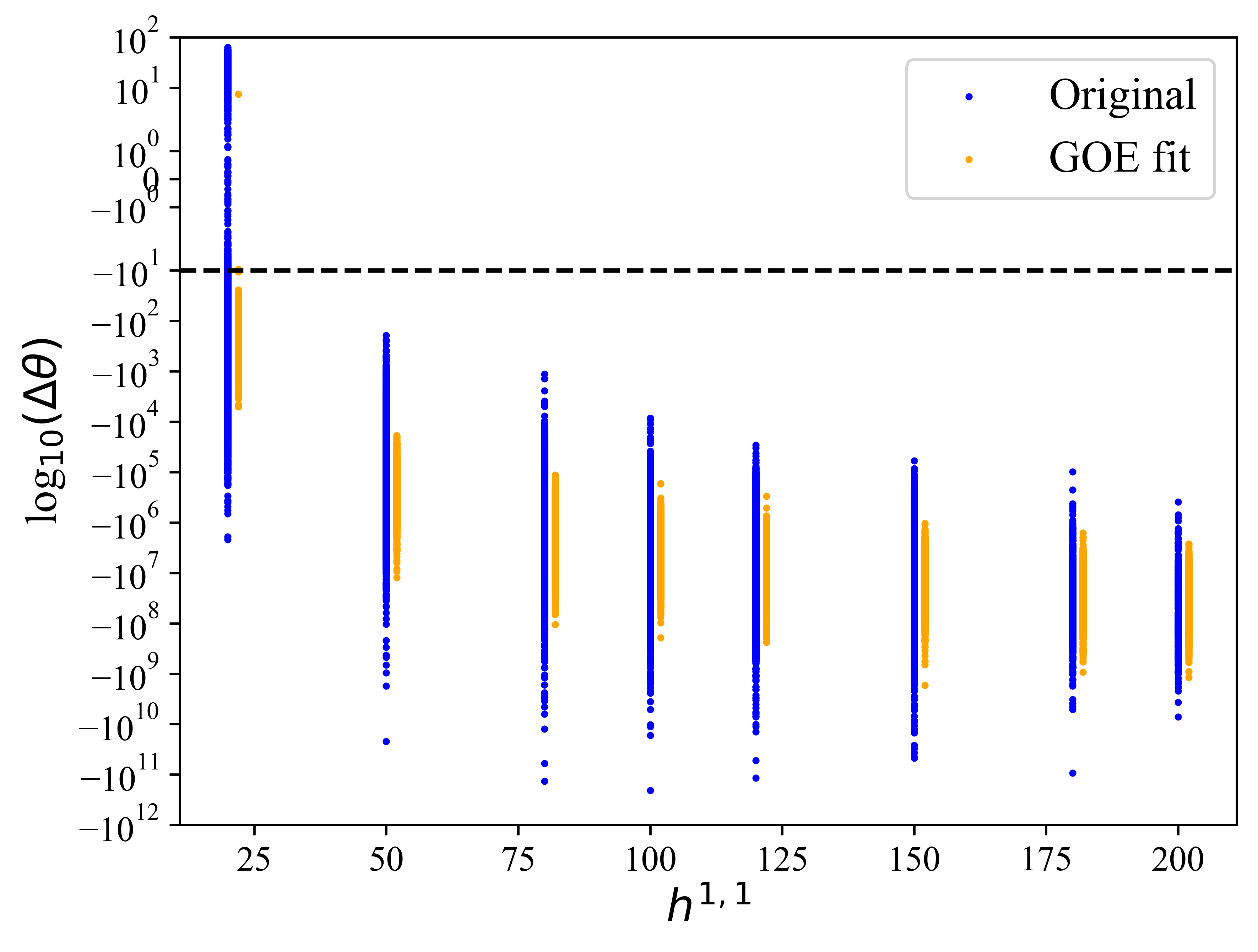}
    \caption{Distributions of $\Delta\theta$ computed using $\{\tau^a\}$ and $\{\tau^a_{\text{fit}}\}$. ``Original'' and ``GOE fit'' always have the same $h^{1,1}$, but are plotted slightly apart for clarity.
    }
    \label{fig:theta_distribution}
\end{figure}

Finally, we demonstrate that our model can accurately approximate the stringy corrections \eqref{eq:Delta_theta} to the QCD $\theta$ angle. We use the approximation method explained at the end of \S\ref{subsec:divapprox}, with $\{\tau^I\}_X$ replaced by $\{\tau^I\}_{\text{fit}}$ as computed in \S\ref{ssec:which_model}. In \S\ref{subsec:divapprox}, we showed that a reliable approximation for the leading PQ-breaking instanton is the Euclidean D3-brane associated to the $h^{1,1}$-largest divisor. We therefore utilize this approximation here, computing \eqref{eq:Delta_theta} with $\Lambda_{\vec{q}_c}$ computed using the $h^{1,1}$-largest divisor volume in the random draw. The resulting modeled distribution of $\Delta\theta$ is shown in Figure \ref{fig:theta_distribution}, along with the actual stringy contributions computed using the geometric data in the explicit ensemble. Here, we can see that the overall scale of the approximated stringy contributions to $\Delta \theta$ match the values from the true ensemble quite well.

\begin{figure}
    \centering
    \includegraphics[width=0.45\linewidth]{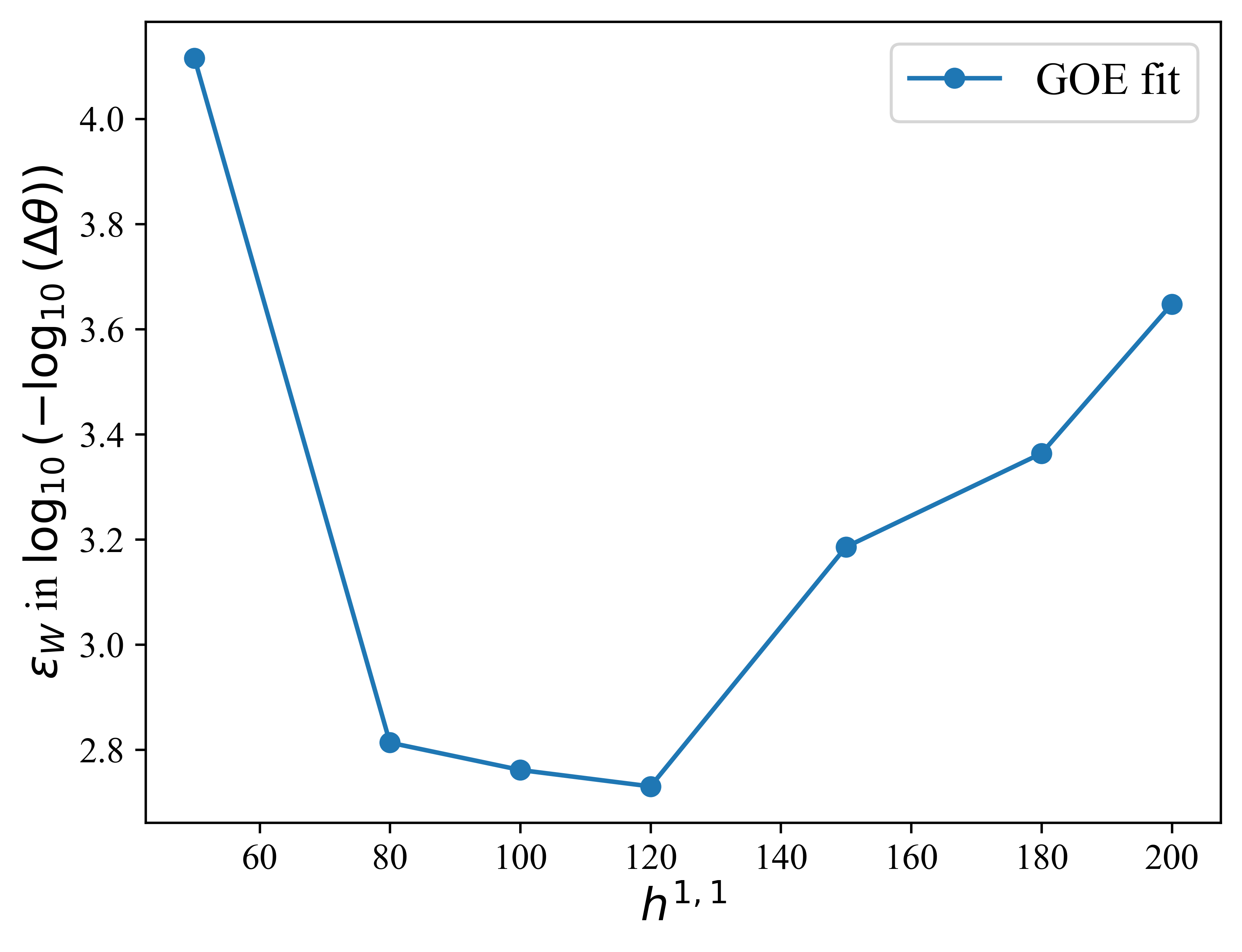}
    \caption{Percentage error between all the original models and all the fitted models with the same $h^{1,1}$.
    }
    \label{fig:fit_theta}
\end{figure}

Finally, we compute $\epsilon_W$ for the distributions of $\Delta \theta$ compared with the actual values in the explicit ensemble as before. Because there is only a single value of $\Delta \theta$ per model, we compare the overall distributions of $\Delta \theta$ for a particular $h^{1,1}$ this time. Fig.~\ref{fig:fit_theta} shows the resulting percentage error in this distance. We observe that the error is low (always under $5 \%$), indicating that the model accurately reproduces string contributions to the QCD $\theta$-angle.

\section{Conclusions} \label{sec:conclusions}

We have analyzed the data defining axion effective theories in compactifications of type IIB string theory on toric hypersurface Calabi-Yau threefolds. This work consists of three main results. Firstly, that these axion EFTs are well approximated using only the data of divisor volumes in the underlying geometries. Secondly, that divisor volume spectra exhibit universal properties across geometries. Finally, that a simple analytic model is sufficient to reproduce axion observables across this landscape. The model is given by \eqref{eq:norm_tau_fit}, \eqref{eq:mean_fit}, and  \eqref{eq:tau_fit}. For reference, we summarize the model below explicitly. The predicted divisor volumes are given by
\begin{align}
    \log(\tau_{\text{fit}}^I) = \hat{\tau}_{\text{fit}}^I \times \langle \log(\vec{\tau})\rangle_{\text{fit}},
    \label{eq:div_fit_conc}
\end{align}
where 
\begin{align}
    \langle \log(\vec{\tau})\rangle_{\text{fit}}=a-b\times(\log_{10}h^{1,1})^{-c},
    \label{eq:mean_fit_conc}
\end{align}
where $a\approx 3.433$, $b \approx 5.404$, and $c \approx 4.050$, and the normalized divisor volumes $\hat{\tau}_{\text{fit}}^I$ are drawn randomly from the distribution
\begin{align}
p_{\text{fit}}(\hat{\tau}) = \frac{\pi}{2} \hat{\tau} \, e^{-\frac{\pi \, \hat{\tau}^2}{4}}.
\end{align}
To then compute axion masses and decay constants with this model, one utilizes the approximations \eqref{eq:m_approx} and  \eqref{eq:f_approx}, repeated here for reference:
\begin{align}
     m_i^{\text{fit}} = \frac{(\tau_{\text{fit}}^i)^{3/4}}{(\tau^{\text{max}}_{\text{fit}})^{3/4}} e^{-\pi \tau_{\text{fit}}^I}, \ \ \ \ \ f_i^{\text{fit}} = \frac{1}{(\tau^{\text{max}}_{\text{fit}})^{3/4} (\tau_{\text{fit}}^i)^{1/4}}.
\end{align}

The motivation for this work is the observation that in repeated studies of axion physics across the string landscape \cite{Gendler:2023hwg, Gendler:2023kjt, Demirtas:2021gsq, Demirtas:2018akl, Halverson:2019cmy, Halverson:2019kna, Mehta:2021pwf, Benabou:2025kgx}, striking patterns were encountered in the phenomenological properties of the axions. In particular, it was found that hierarchies in divisor volumes catalyzed dependence of axion decay constants on $N$, the number of axions. The main goal of this work was to make progress in quantifying these hierarchies. Indeed, we found that divisor volume spectra can be reliably modeled as the distribution \eqref{eq:div_fit_conc}, with a mean that increases  $h^{1,1}$ as \eqref{eq:mean_fit_conc}. Together, these approximations serve as a model of divisor volume distributions as a function of $h^{1,1}$.

The utility of this model is that it can serve as a means to approximate phenomenological characteristics of axions in this landscape, without the need for computationally expensive scans, or even any knowledge of toric geometry or string theory. A key step in establishing this was to show that the data of divisor volumes alone are sufficient to reproduce the observables of axion EFTs from string compactifications. This fact was our main finding in \S\ref{sec:axiondivs}. There, we presented a set of approximations for axion masses, decay constants, and contributions to the QCD $\theta$-angle that depend only on a given set of divisor volumes. Using statistical tests (the $1$-Wasserstein distance and adjusted $r^2$ test), we showed that these expressions provide good estimates for these quantities. 

In \S\ref{sec:universal}, we showed that distributions of normalized divisor volumes (specifically, logarithms of divisor volumes normalized by the mean log divisor volume of a given geometry---see \eqref{eq:normdivs}) follow similar distributions, regardless of the point in moduli space, the specific geometry chosen, or even the Hodge number $h^{1,1}$ of the Calabi-Yau threefold. We again corroborated this fact with statistical tests, finding that divisor volume distributions become more similar as $h^{1,1}$ increases. We noted the emergence of a second peak in the distribution at moderate $h^{1,1}$, which we found was attributable to the presence of three underlying sub-distributions, corresponding to vertex, edge, and face divisors (defined explicitly in \S\ref{subsec:batyrev}). These sub-distributions also exhibit universal features, which we again checked using statistical tests.

Armed with an underlying distribution for normalized divisor volumes, in \S\ref{sec:model}, we presented an analytical approximation for this distribution, which, when combined with a regression on the value of the mean divisor volume as a function of $h^{1,1}$, serves as a simple model for divisor volume spectra across the Kreuzer-Skarke landscape. The best fit model to normalized divisor volumes was the level spacing distribution of the Gaussian Orthogonal Ensemble, suggesting a tantalizing connection between Calabi-Yau geometry and random matrix theory. We showed that our model can reliably reproduce axion masses and decay constants, as well as approximate stringy contributions to the QCD $\theta$-angle.

This work serves as a first step towards understanding what geometric features drive patterns in axion physics from string theory. Here, we have quantified how hierarchies in divisor volumes change throughout a diverse ensemble of Calabi-Yau threefolds generated as hypersurfaces in toric varieties. An important avenue for future work is to understand how these distributions change in classes of geometries beyond those studied here, e.g. in CICY Calabi-Yau threefolds, Calabi-Yau fourfolds, and in non-Calabi-Yau geometries. Additionally, it is crucial to understand what the underlying geometric features that dictate these distributions are. In particular, we found in \S\ref{sec:model} that a good approximation for the spectrum of normalized divisor volumes is the level-spacing distribution of the Gaussian Orthogonal Ensemble. Significant effort has been put into connecting Calabi-Yau geometry and the string theory landscape with random matrix theory (see e.g. \cite{Afkhami-Jeddi:2021qkf, Ferrari:2011we, Long:2014fba, Bachlechner:2014rqa, Bachlechner:2019vcb, Marsh:2011aa, Dijkgraaf:2002fc, Easther:2005zr, DiFrancesco:1993cyw}). A worthwhile investigation is to understand whether our model hints at such a connection.

Finally, a promising extension of this work is to use our findings on statistics of divisor volume distributions in order to investigate statistics of vacua in the string landscape, along the lines of \cite{Douglas:2003um, Denef:2004ze}. In particular, it would be interesting to understand whether divisor volume distributions that are significantly different from the ones we present here (as is necessary, for example, in the KKLT or LVS \cite{Balasubramanian:2005zx}  scenarios) occur with probabilities dictated by our model.

\section*{Acknowledgments}

We are grateful to David J. E. Marsh, Liam McAllister, Jakob Moritz, Matthew Reece, John Stout, and Cumrun Vafa for useful discussions and feedback. We thank Liam McAllister, Jakob Moritz,  and Matthew Reece for comments on a draft. The work of JC was supported in part by the DOE Grant DE-SC-0013607. The work of NG
was supported in part by a grant from the Simons
Foundation (602883,CV) and the DellaPietra Foundation.

\bibliographystyle{JHEP}
\bibliography{refs}

\end{document}